\documentclass[11pt,a4paper]{article}
\pdfoutput=1
\usepackage{jheppub}

\usepackage{multirow, graphicx,amssymb,url,mathrsfs,amsmath}
\usepackage{wrapfig,boxedminipage,setspace,subfigure,epsfig}
\usepackage{amsxtra,amstext,latexsym,dsfont,amsfonts}
\usepackage{color,eucal}
\usepackage[dvipsnames]{xcolor}
\usepackage{float}
\usepackage{slashed,comment}

\usepackage{tabularx, array}
\newcolumntype{L}[1]{>{\raggedright\arraybackslash}p{#1}}
\newcolumntype{C}[1]{>{\centering\arraybackslash}p{#1}}
\newcolumntype{R}[1]{>{\raggedleft\arraybackslash}p{#1}}

\newcommand{\td}{\text{d}}

\newcommand{\ra}{\rightarrow}

\newcommand{\dd}{\mathrm{d}}

\newcommand{\csch}{\text{csch}}
\newcommand{\sech}{\text{sech}}

\title{Pole-Skipping in Rotating BTZ Black Holes}

\author[a,b]{Hyun-Sik Jeong,}
\author[c]{Chang-Woo Ji,}
\author[c,d]{and Keun-Young Kim}

\emailAdd{hyunsik.jeong@uam.es}
\emailAdd{physianoji@gm.gist.ac.kr}
\emailAdd{fortoe@gist.ac.kr}

\affiliation[a]{Instituto de F\'isica Te\'orica UAM/CSIC, Calle Nicol\'as Cabrera 13-15, 28049 Madrid, Spain}
\affiliation[b]{Departamento de F\'isica Te\'orica, Universidad Aut{\'o}noma de Madrid, Campus de Cantoblanco, 28049 Madrid, Spain}
\affiliation[c]{Department of Physics and Photon Science, Gwangju Institute of Science and Technology, \\
123 Cheomdan-gwagiro, Gwangju 61005, Korea}
\affiliation[d]{Research Center for Photon Science Technology, Gwangju Institute of Science and Technology, \\
123 Cheomdan-gwagiro, Gwangju 61005, Korea}

\preprint{IFT-UAM/CSIC-23-56}

\abstract{
Motivated by the connection between pole-skipping phenomena of two point functions and four point out-of-time-order correlators, we study the pole-skipping phenomena for rotating BTZ black holes. In particular, we investigate the effect of rotations on the pole-skipping point for various fields with spin $s = 1/2, 1, 2/3$, extending the previous research for $s=0, 2$. We derive an analytic full tower of the pole-skipping points of fermionic ($s=1/2$) and vector ($s=1$) fields by the exact holographic Green's functions. For the \textit{non-extremal} black hole, the leading pole-skipping frequency is $\omega_{\text{leading}}=2\pi i T_h {(s-1+\nu \Omega)}/{(1-\Omega^2)}$ where $T_h$ is the temperature, $\Omega$ the rotation, and $\nu:=(\Delta_+ - \Delta_-)/2$, the difference of conformal dimensions ($\Delta_{\pm}$). These are confirmed by another independent method: the near-horizon analysis. For the \textit{extremal} black hole, we find that the leading pole-skipping frequency can occur at  $\omega_{\text{leading}}^{\text{extremal}}=-2\pi i T_R {(s+1)}$ only when $\nu = s+1$, where $T_R$ is the temperature of the right moving mode. It is non-trivial because it cannot be achieved by simply taking the extreme limit ($T_h\rightarrow 0\,, \Omega\rightarrow 1$) of the non-extremal black hole result.
}

\begin{document}
\maketitle

%
\section{Introduction}\label{}
Retarded Green's functions, denoted as $G^R$, play a fundamental role in the field of physics by providing essential insights into the behavior and properties of various physical systems. Notably, the exploration of the AdS/CFT duality or holographic duality~\cite{Maldacena:1997re,Witten:1998qj,Witten:1998zw,Gubser:1998bc} has uncovered a remarkable and \textit{universal} characteristic inherent in Green's functions, commonly referred to as pole-skipping~\cite{Grozdanov:2017ajz,Blake:2018leo}.

The Green's function of the dual field theory in holography is typically associated with the fluctuations in the bulk. Perturbation problems in the bulk often involve different fields, including scalar fields, Dirac fields, Maxwell fields, and gravitational fields. However, when studying the dual Green's functions in the complex momentum space ($\omega, k$), particularly at pole-skipping points ($\omega_{*}, k_{*}$), the uniqueness of Green's functions becomes uncertain. For instance, near the pole-skipping point, the Green's function can be expressed as 
\begin{align}\label{VFEQIDP}
\begin{split}
G^{R}(\omega_{*}, k_{*}) \approx \frac{0}{0}\,, \qquad G^{R}(\omega_{*}+\delta \omega, k_{*}+\delta k) \approx \frac{\delta \omega + \delta k}{\delta \omega - \delta k} \,,
\end{split}
\end{align}
where the Green's function becomes non-unique and its behavior is dependent on the slope $\delta \omega/\delta k$ as it approaches the pole-skipping point. From a bulk perspective, the appearance of pole-skipping arises due to the non-uniqueness of the bulk solution, especially due to the existence of ``two" independent ingoing solutions at the black hole horizon.

The phenomenon of pole-skipping was initially discovered in the context of holographic chaos~\cite{Grozdanov:2017ajz,Blake:2018leo}, where the presence of pole-skipping points in the energy density two-point function is closely related to the Lyapunov exponent (and butterfly velocity) that govern the behavior of out-of-time-ordered correlation functions. This intriguing pole-skipping phenomenon arises as a prediction of hydrodynamic effective theories that describe maximally chaotic systems~\cite{Blake:2017ris,Blake:2021wqj}, and it has been observed to be a generic feature of holographic theories dual to \textit{planar} black holes~\cite{Blake:2018leo,Jeong:2021zhz}.\footnote{Further insights into pole-skipping beyond maximal chaos have been proposed in~\cite{Choi:2020tdj}, and investigations of pole-skipping in conformal field theories can be found in~\cite{Haehl:2018izb,Jensen:2019cmr,Das:2019tga,Haehl:2019eae,Ramirez:2020qer}.
The similar phenomenon of pole-skipping has also been observed in quantum mechanics problems~\cite{Natsuume:2021fhn}, indicating the relevance of this phenomenon within a broader research context.}

Since the discovery of the connection to quantum chaos, significant attention has been focused on investigating the mathematical foundations of the pole-skipping phenomena. The striking progress in this research area includes the establishment of the \textit{universality} of pole-skipping points, as demonstrated in studies such as \cite{Grozdanov:2019uhi,Blake:2019otz,Natsuume:2019xcy,Natsuume:2019sfp,Natsuume:2019vcv,Wu:2019esr,Ceplak:2019ymw,Ahn:2019rnq,Ahn:2020bks,Abbasi:2020ykq,Liu:2020yaf,Ramirez:2020qer,Ahn:2020baf,Natsuume:2020snz,Kim:2020url,Sil:2020jhr,Ceplak:2021efc,Kim:2021hqy,Jeong:2021zhz,Natsuume:2021fhn,Blake:2021hjj,Kim:2021xdz,Jeong:2022luo,Wang:2022xoc,Wang:2022mcq,Amano:2022mlu,Yuan:2023tft,Baishya:2023nsb,Grozdanov:2023txs,Natsuume:2023lzy}.

There exists a remarkable universality for the pole-skipping points, particularly at their leading expression, which includes Matsubara frequencies:
\begin{equation} \label{UNFOR1}
    \omega_{s}=2\pi i T_h \left( s-1 \right)\,,
\end{equation}
where $s$ denotes the spin of the bulk fields and $T_h$ the black hole temperature. For instance, the scalar, Dirac, Maxwell (or vector), and gravitational spin-2 field have the pole-skipping points as
\begin{equation} \label{}
    \omega_{0} = -2\pi i T_h \,, \quad  \omega_{1/2} = -\pi i T_h \,, \quad  \omega_{1} = 0 \,, \quad \omega_2 = + 2 \pi i T_h\,,
\end{equation}
where the case of the vector field, $\omega_{1}$, corresponds to the hydrodynamic pole.
In the context of the gravitational sound mode (spin-2 field), the corresponding pole-skipping point $\omega_2$ has been proposed that it is associated with the many-body quantum chaos~\cite{Shenker:2013pqa,Roberts:2014isa,Roberts:2014ifa,Shenker:2014cwa,Maldacena:2015waa}.\\

More recent investigations have explored the relationship between chaos and energy dynamics in \textit{rotating} black holes in the context of pole-skipping phenomena. A notable case is the rotating BTZ black hole (three-dimensional gravity theories), which is dual to a two-dimensional conformal field theory featuring a chemical potential for rotation. Detailed examinations of this scenario were conducted, for instance, in \cite{Liu:2020yaf}.\footnote{See also \cite{Craps:2021bmz,Jahnke:2019gxr} for the additional derivations of the form of the out-of-time-ordered correlation function.} Additionally, the interplay of chaos and pole-skipping in the field theory dual to a higher-dimensional rotating Kerr-AdS black hole was recently investigated in \cite{Blake:2021hjj,Amano:2022mlu}.\footnote{In \cite{Blake:2021hjj,Amano:2022mlu}, they derived a partial differential equation that governs the angular profile of gravitational shock waves, which is essential for calculating out-of-time-ordered correlation functions.}

When taking into account finite rotation, an intriguing consequence arises: the universal pole-skipping point can be extended to incorporate rotational effects. Previous studies, such as \cite{Liu:2020yaf,Blake:2021hjj,Amano:2022mlu}, have revealed that $\omega_2$ can be generalized as follows:
\begin{equation} \label{UNFOR2}
    \omega_{2}=\frac{2\pi i T_h}{1-\Omega^2} \left( 1 + \nu \Omega \right)\,.
\end{equation}
Here, $\Omega$ represents the finite rotation and $\nu$ is associated with the conformal dimension of the dual boundary operator. 
In addition, through the analysis of scalar field fluctuations, \cite{Natsuume:2020snz} revealed that the value of $\omega_0$ can also be further extended to $\omega_{0}=-\frac{2\pi i T_h}{1-\Omega^2} \left( 1 - \nu \Omega \right)$.\\

In this paper, we focus our investigation on the Dirac ($\omega_{1/2}$) and vector ($\omega_{1}$) fields, with the objective of extending \eqref{UNFOR1} to derive the more general universal pole-skipping point. This generalization incorporates both the spin information, represented by $s$, and the rotation parameter $\Omega$.

Additionally, as a result of considering finite rotation, we also explore the pole-skipping point in the extreme limit ($T_h\rightarrow 0\,, \Omega\rightarrow 1$). As we will show in the main context, it is imperative to note that this limit can only be approached when the analytic Green's function is applicable. Furthermore, in the context of extremal black holes, we will reveal an additional universal relationship associated with the pole-skipping point.\\

The structure of this paper is as follows.
In section {2}, we provide a quick review of the black hole background of our interest, specifically the rotating BTZ black holes. Building upon the BTZ black holes introduced in section {2}, in section {3}, we focus on the fermionic fluctuation field and derive the analytic Green's function. In addition, we perform the systematic analysis of the pole-skipping points from such Green's function. Likewise, in section {4}, we investigate the exact Green's function of massive vector fields, and examine the analytic structure of pole-skipping points in presence of rotations. 
Section {5} is devoted to conclusions.

%
\section{The rotating BTZ black holes}\label{sec:The rotating BTZ black holes}

Let us first briefly review the rotating BTZ black hole~\cite{Banados:1992wn,Banados:1992gq} whose metic reads in terms of Schwarzschild coordinates ($t,r,\varphi$) as
\begin{align}\label{eq:metric}
\begin{split}
    {\td s}^2 & =-f(r) {\td t}^2+\frac{{\td r}^2}{f(r)}+r^2 \left(\td\varphi-\frac{J}{2r^2}\td t \right)^2\,,
\end{split}
\end{align}
where $\varphi$ is periodic with $2\pi$ and $J$ angular momentum given in \eqref{MANDJ}. 
The emblackening factor $f(r)$ can be found as the solution of Einstein gravity 
\begin{align}\label{EMF}
\begin{split}
    f(r) = r^2 - M + \frac{J^2}{4 r^2} = \frac{(r^2-r_+^2)(r^2-r_-^2)}{r^2}\,,
\end{split}
\end{align}
where the black hole mass $M$ and angular momentum $J$ are 
\begin{align}\label{MANDJ}
\begin{split}
    M=r_+^2+r_-^2 \,, \qquad J =2 r_+ r_- \,.
\end{split}
\end{align}

Using the emblackening factor \eqref{EMF}, one can also find the Hawking temperature $T_h$ and other thermodynamic quantities such as the entropy density $S$:
\begin{align}
\begin{split}
T_h = \frac{f'(r_+)}{4\pi} = \frac{r_+^2-r_-^2}{2\pi \, r_+} \,, \qquad S = 4\pi r_+ \,, \qquad J = 2 r_+^2 \Omega \,,
\end{split}
\end{align}
where the angular velocity (or angular momentum potential) $\Omega$ is
\begin{align}\label{MANDJ2}
\begin{split}
\Omega  = \frac{r_-}{r_+} \,.
\end{split}
\end{align}
where $0 \leq \Omega \leq 1$.
Note that the thermodynamic quantities \eqref{MANDJ}-\eqref{MANDJ2} satisfy the first law of thermodynamics 
\begin{align}\label{}
\begin{split}
\dd M \,=\, T_h \, \dd S \,+\, \Omega \, \dd J \,.
\end{split}
\end{align}

Furthermore, using the angular velocity $\Omega$ \eqref{MANDJ2}, one may also introduce temperatures of the left/right moving modes in a two dimensional CFT as
\begin{align}\label{eq:TLTR OmegaT}
\begin{split}
     T_L  = \frac{T_h}{1+\Omega} = \frac{r_+ - r_-}{2 \pi}\,,\qquad
    T_R  = \frac{T_h}{1-\Omega} = \frac{r_+ + r_-}{2 \pi}\,.
\end{split}
\end{align}
Following the standard convention we choose $T_L\rightarrow0$ for the \textit{extreme limit} in this paper.\\

In following sections, we will study the fluctuation fields (fermionic field $\psi$ and gauge field $A$) on the background geometry \eqref{eq:metric}. In particular, we aim to solve the corresponding fluctuation equations of motion and obtain the Green's function analytically. 
For this purpose, it is convenient to introduce another coordinate $(T, \rho, X)$
\begin{equation}\label{eq:metric TrhoX}
    {\td s}^2 = -\sinh^2{\rho} \,\, {\td T}^2 + {\td \rho}^2 + \cosh^2{\rho} \,\, {\td X}^2\,,
\end{equation}
where the coordinate transformation from \eqref{eq:metric} to \eqref{eq:metric TrhoX} is given
\begin{equation} \label{eq:TorhoTX}
    r^2 = r_+^2 \cosh^2{\rho} - r_-^2 \sinh^2{\rho}\,,\quad
    T+X=(r_+-r_-)(t+\varphi)\,,\quad
    T-X=(r_+ + r_-)(t-\varphi)\,.
\end{equation}
Note that the AdS boundary in the Schwarzschild coordinates \eqref{eq:metric}, $r \rightarrow \infty$, becomes $\rho \rightarrow \infty$ in a new coordinate \eqref{eq:metric TrhoX}.

%
\section{Fermionic fields ($s=1/2$)}\label{sec:Fermionic fields}

In this section, we study the pole-skipping point of fermionic field in the presence of rotations.
For this purpose, we will follow closely the presentation of \cite{Iqbal:2009fd} in order to obtain the Green's function.
Furthermore, we consider the pole-skipping point from the Green's function not only for the non-extremal case, but also for the extremal case. As we will show shortly, the non-extremal case is consistent with \cite{Ceplak:2019ymw}, while the extremal case has not been reported in literature yet.

%
\subsection{Dirac equations of motion}
In this section, we study the fermionic pole-skipping points in the presence of the rotation. For this purpose, we start with a Dirac equation written in terms of a Dirac spinor field $\psi$ with the fermion mass $m_f$:
\begin{equation} \label{eq:DiracEqn}
    \left( \Gamma^M D_M - m_f \right) \psi = 0\,
\end{equation}
where $\Gamma^{M}$ is the gamma matrix and $D_M$ the covariant derivative.
In this paper, we use the indices $M, N=(T,\rho,X)$ for the bulk spacetime and $a, b=(\underline{T},\underline{\rho},\underline{X})$ for the tangent spacetime where they are associated by the veinbein $e^a_M$ as 
\begin{equation} \label{MEVER}
    g_{MN} = \eta_{ab} \, e^a_M \, e^b_N \,, \qquad \eta_{ab} = (-1,1,1) \,.
\end{equation}

In terms of the veinbein, $\Gamma^{M}$ and $D_M$ can be expressed as
\begin{equation} \label{}
    \Gamma^{M} = \Gamma^a e^M_a \,, \qquad D_M = \partial_M + \frac{1}{4} \left( \omega_{ab} \right)_M \Gamma^{ab} \,,
\end{equation}
where the spin connection $\left( \omega_{ab} \right)_M$ and $\Gamma^{ab}$ are determined by  
\begin{equation} \label{MEVERL}
 {
 (\omega^a_b)_M = e^a_N \, e^Q_b \, \Gamma^N_{MQ} - e^Q_b \, \partial_{M} e^a_Q \,,
 \qquad \Gamma^{ab} \, e^M_a \, e^N_b = \frac{1}{2} \left[\Gamma^M, \Gamma^N\right]} \,,
\end{equation}
with the Christoffel symbols $\Gamma^N_{MQ}$.

Given all the relations \eqref{MEVER}-\eqref{MEVERL}, in order for solving the Dirac equation \eqref{eq:DiracEqn}, we need to specify two things: i) the inverse vielbein $e^M_a$; ii) the gamma matrix $\Gamma_a$.
In what follows, given the metric $g_{MN}$ \eqref{eq:metric TrhoX}, we choose the diagonal inverse vielbein as
\begin{equation} \label{eq:VielBein}
    e^{T}_{\underline{T}} = \csch{\rho} \,, \qquad e^{\rho}_{\underline{\rho}} = 1 \,, \qquad e^{X}_{\underline{X}} = \sech{\rho} \,,
\end{equation}
together with the gamma matrix
\begin{equation}\label{eq:Gamma Matrices}
    \Gamma^{\underline{T}}=\begin{pmatrix}
    0 & 1\\
    -1 & 0
    \end{pmatrix}\,,\qquad
    \Gamma^{\underline{\rho}}=\begin{pmatrix}
    1 & 0\\
    0 & -1
    \end{pmatrix}\,,\qquad
    \Gamma^{\underline{X}}=\begin{pmatrix}
    0 & 1\\
    1 & 0
    \end{pmatrix}\,.
\end{equation}

\paragraph{Dirac equations in the Fourier space.}
Furthermore, introducing the Dirac spinor $\psi$ in the Fourier space\footnote{Note that $\psi_{\pm}$ in \eqref{eq:fermionic field ansatz} are the eigenvectors of the gamma matrix $\Gamma^{\underline{\rho}}$, i.e., $\Gamma^{\underline{\rho}} \, \psi_{\pm}=\pm \psi_{\pm}$.}
\begin{equation} \label{eq:fermionic field ansatz}
\psi=
    \begin{pmatrix}
    \psi_+(\rho) \\ \psi_-(\rho)
    \end{pmatrix}
    e^{-i k_T T \,+\, i k_X X}\,,
\end{equation}
we find the Dirac equations of motion \eqref{eq:DiracEqn} as
\begin{equation}\label{eq:DiracEqnfour}
    \left[ \Gamma^{\underline{\rho}} \left( \partial_\rho + \frac{1}{2} \left( \frac{\cosh{\rho}}{\sinh{\rho}} + \frac{\sinh{\rho}}{\cosh{\rho}} \right) \right) + i \left( \frac{k_X \Gamma^{\underline{X}}}{\cosh{\rho}} - \frac{k_T \Gamma^{\underline{T}}}{\sinh{\rho}} \right) - m_f \right] \psi = 0\,.
\end{equation}
In order to solve this equation analytically, it is convenient to use the ansatz
\begin{equation}\label{eq:psiTochi}
\psi_\pm(z)=\sqrt{\frac{\left(1\pm\sqrt{z}\right)\sqrt{1-z}}{\sqrt{z}}}\left( \chi_1(z) \pm \chi_2(z) \right)\,, \qquad z=\tanh^2{\rho}\,,
\end{equation}
where the coordinate $\rho$ is replaced by $z$. Note that in this $z$-coordinate, the AdS boundary is located at $z\rightarrow1$.

Within the ansatz \eqref{eq:psiTochi}, the Dirac equations \eqref{eq:DiracEqnfour} become
\begin{align}\label{eq:Dirac Eqn chi12 a}
\begin{split}
    2(1-z)\sqrt{z} \, \partial_z \chi_1 - i\left( \frac{k_T}{\sqrt{z}} + k_X \sqrt{z} \right) \chi_1 &= \left( m_f-\frac{1}{2}+i(k_T+k_X) \right)\chi_2\,, \\
    2(1-z)\sqrt{z} \, \partial_z \chi_2 + i\left( \frac{k_T}{\sqrt{z}} + k_X \sqrt{z} \right) \chi_2 &= \left( m_f-\frac{1}{2}-i(k_T+k_X) \right)\chi_1\,,
\end{split}
\end{align}
which allow the \textit{analytic} incoming solutions in terms of the hypergeometric function ${}_2 F_{1}$ as
\begin{align}\label{eq:chiSol}
\begin{split}
    \chi_1(z) &\,=\, \left( \frac{a-c}{c} \right) \, z^{\alpha+\frac{1}{2}} \, (1-z)^\beta \, {}_2 F_{1}(a,b+1;\,c+1;\,z) \,, \\
    \chi_2(z) &\,=\,  z^\alpha \, (1-z)^\beta \, {}_2 F_{1}(a,b;\,c;\,z)\,,
\end{split}
\end{align}
where
\begin{align}\label{abcre}
\begin{split}
a &= \frac{1}{2} \left( m_f+\frac{1}{2} \right) - \frac{i}{2} \left(k_T -  k_X\right)\,, \qquad 
b = \frac{1}{2} \left( m_f-\frac{1}{2} \right) - \frac{i}{2} \left( k_T + k_X \right)\,,\\
c &= \frac{1}{2}+a+b-m_f  = \frac{1}{2} - i k_T  \,,
\end{split}
\end{align}
and
\begin{equation}\label{ABCOEFF}
    \alpha = - \frac{i k_T}{2}\,, \qquad \beta = - \frac{1}{4} + \frac{m_f}{2}\,.
\end{equation}
Therefore, using the ansatz \eqref{eq:psiTochi} together with \eqref{eq:chiSol}-\eqref{ABCOEFF}, one can find the analytic fermionic field $\psi_\pm(z)$.
%

%
\subsection{Exact fermionic Green's function} \label{Exact fermionic Green's function}

\subsubsection{Holographic dictionary for fermionic Green's function}
Next, using the analytic solution $\psi_\pm(z)$ obtained above, we study the fermionic Green's function. Let us first review how to read such a fermionic boundary correlator.

\paragraph{Non half-integer case.}
According to the holographic dictionary, the Green's function are related to the AdS boundary behavior of fermionic fields $\psi_\pm(z)$.
For instance, one can find that the equation of motion \eqref{eq:DiracEqnfour} within a $z$-coordinate \eqref{eq:psiTochi}, $z=\tanh^2{\rho}$, produces the following asymptotic behavior near the AdS boundary ($z\rightarrow1$)
\begin{align}\label{eq:Non-Integer boundary behavior}
\begin{split}
    \psi_+ &\,\approx\, \mathcal{A} (1-z)^{\frac{\Delta_{-}}{2}} \,+\, \mathcal{B} (1-z)^{\frac{1+\Delta_{+}}{2}}  \,+\, \cdots  \,, \\
    \psi_- &\,\approx\, \mathcal{C} (1-z)^{\frac{1+\Delta_{-}}{2}} \,+\, \mathcal{D} (1-z)^{\frac{\Delta_{+}}{2}} \,+\, \cdots \,,
\end{split}
\end{align}
where $\Delta_{\pm}$ is the conformal dimension
\begin{equation}\label{SDR}
    \Delta_{\pm} = 1\pm m_f \,.
\end{equation}
Here, $\mathcal{A}$ and $\mathcal{D}$ are independent free parameters in which the remaining coefficients, $\mathcal{B}$ and $\mathcal{C}$, are determined by
\begin{equation}\label{CONDBC}
    \mathcal{B}=\frac{i(k_X-k_T)}{2m_f+1} \mathcal{D}\,, \qquad
    \mathcal{C}=\frac{i(k_X+k_T)}{2m_f-1} \mathcal{A}\,.
\end{equation}
Using the obtained analytic solution $\eqref{eq:chiSol}$, $\mathcal{A}$ and $\mathcal{D}$ can be found, for instance \eqref{FRAD}.
One can also notice that the relation \eqref{CONDBC} may not be well defined for the case $m_f=\pm1/2$. We will discuss this point later.

\paragraph{Half-integer case.}
It would also be convenient to introduce a new parameter $\nu$ composed of the conformal dimension \eqref{SDR}
\begin{equation}\label{nudef}
  \nu := \frac{\Delta_+ - \Delta_-}{2} = m_f  \,.
\end{equation}
In particular, using $\nu$, we can rewrite the boundary expansion \eqref{eq:Non-Integer boundary behavior} as 
\begin{align}\label{BCRE22}
\begin{split}
    \psi_+ &\,\approx\, \mathcal{A} (1-z)^{\frac{\Delta_{-}}{2}} \,+\, \mathcal{B} (1-z)^{\frac{\Delta_{-}}{2} \,+\, \left(\nu + \frac{1}{2}\right)}  \,+\, \cdots  \,, \\
    \psi_- &\,\approx\, \mathcal{C} (1-z)^{\frac{\Delta_{+}}{2} \,-\, \left(\nu - \frac{1}{2}\right)} \,+\, \mathcal{D} (1-z)^{\frac{\Delta_{+}}{2}} \,+\, \cdots \,.
\end{split}
\end{align}
In other words, in \eqref{eq:Non-Integer boundary behavior}, it is intrinsically assumed
that $\nu$ (or equivalently the mass $m_f$) is not a half-integer:
\begin{align}\label{}
\begin{split}
\nu \pm \frac{1}{2} \,\notin\, \mathbb{Z}^{+}\,.
\end{split}
\end{align}
In principle, it is also possible to consider the ``negative" half-integer. However, it is enough to consider the positive half-integer case for our purpose: see the description around \eqref{nurange2}.

Therefore, for the case of half-integer $\nu$, $\nu \pm \frac{1}{2} \in \mathbb{Z}^{+}$, such as
\begin{align}\label{nurange}
\begin{split}
\nu =   \frac{3}{2}\,,\, \frac{5}{2}\,,\, \frac{7}{2} \,,\, \cdots\,,
\end{split}
\end{align}
one can find the different boundary expansion
\begin{equation}\label{eq:Integer boundary behavior}
\begin{aligned}
    \psi_+  &\,\sim\,  \bar{\mathcal{A}} (1-z)^{\frac{\Delta_{-}}{2}} \left\{1 + \cdots +\mathcal{O}\left( (1-z)^{\nu+\frac{1}{2}} \ln{(1-z)}\right)  \right\} \,+\, \bar{\mathcal{B}} (1-z)^{\frac{1+\Delta_{+}}{2}} \left\{1+\cdots\right\} \,,
    \\
    \psi_-  &\,\sim\, \bar{\mathcal{C}} (1-z)^{\frac{1+\Delta_{-}}{2}} \left\{1 +\cdots+ \mathcal{O}\left( (1-z)^{\nu-\frac{1}{2}} \ln{(1-z)}\right) \right\} \,+\, \bar{\mathcal{D}} (1-z)^{\frac{\Delta_{+}}{2}} \left\{1+\cdots\right\}   \,,
\end{aligned}
\end{equation}
where $\bar{\mathcal{A}}$ and $\bar{\mathcal{D}}$ are still the independent free parameters together with the same relation \eqref{CONDBC}, i.e.,
\begin{equation}\label{CONDBC2}
    \bar{\mathcal{B}}=\frac{i(k_X-k_T)}{2m_f+1} \bar{\mathcal{D}}\,, \qquad
    \bar{\mathcal{C}}=\frac{i(k_X+k_T)}{2m_f-1} \bar{\mathcal{A}}\,.
\end{equation}
Similar to the non half-integer case, using the obtained analytic solution $\eqref{eq:chiSol}$, one can also find the analytic expression for $\bar{\mathcal{A}}$ and $\bar{\mathcal{D}}$ as \eqref{FRAD2}.

\paragraph{Fermionic Green's function in holography.}
According to the holographic principle, the fermionic Green's function, $\tilde{G}^R$, is given by the ratio between two independent parameters $\mathcal{D}$ and $\mathcal{A}$ as  
\begin{equation}\label{eq:Gtilde}
    \tilde{G}^R = i \frac{\mathcal{D}}{\mathcal{A}}  \quad \text{or} \quad i \frac{\bar{\mathcal{D}}}{\bar{\mathcal{A}}} \,,
\end{equation}
where the former one is for the non half-integer case, while the later is for the half-integer case.
Strictly speaking, the Green's function in \eqref{eq:Gtilde} corresponds to the case of the standard quantization where $\mathcal{A}$ (or $\bar{\mathcal{A}}$) is interpreted as the source and $\mathcal{D}$ (or $\bar{\mathcal{D}}$) is the corresponding vev. 

In principle, it is also possible to consider the other quantization, alternative quantization, by replacing $\mathcal{A} \leftrightarrow \mathcal{D}$ or $\bar{\mathcal{A}}\leftrightarrow \bar{\mathcal{D}}$.
However, in this paper, we only consider the case of standard quantization \eqref{eq:Gtilde} for our own purpose: notice that the structure of the pole-skipping, $\tilde{G}^{R}\sim 0/0$, is independent of the type of a quantization.

Equivalently, considering a standard quantization implies that we only focus on the positive $\nu$, for instance, one can notice that the role of $\mathcal{A}$ and $\mathcal{D}$ in
\eqref{BCRE22} can be exchanged when the sign of $\nu$ in \eqref{nudef} is reversed.

Therefore, in this paper, we only choose $\nu \geq 0$ hereafter, i.e.,
\begin{align}\label{nurange2}
\begin{split}
\nu = 
\begin{cases}
       \,\, \textrm{Otherwise} \qquad\,\, (\textrm{Non half-integer case}) \,, \\
       \,\, \frac{3}{2}\,,\, \frac{5}{2} \,,\, \frac{7}{2} \,,\, \cdots\, \quad (\textrm{Half-integer case}) \,.
\end{cases}
\end{split}
\end{align}
Mathematically, it is also possible to consider $\nu=m_f=1/2$. However, this special case may produce the ill-defined relation \eqref{CONDBC} (or \eqref{CONDBC2}) and requires the separate calculations for the Green's function: for instance, see~\cite{Iqbal:2009fd}.
Furthermore, it is also worth noting that this special case can also give rise to the subtlety in the boundary expansion~\eqref{BCRE22}.

In this study, we do not explore this special case for simplicity. Note that the special case can also appear even in the scalar field case~\cite{Natsuume:2020snz} with $\nu=0$. Nevertheless, it is shown \cite{Natsuume:2020snz} that the pole-skipping cannot occur in this case. We suspect that this may also be the case even for other fields. We leave this subject as future work.\\

\paragraph{Fermionic Green's function in $(t,r,\varphi)$ coordinate.}
Following the coordinate transformation \eqref{eq:TorhoTX}, it is shown \cite{Iqbal:2009fd} that one can restore the Green's function in the original coordinate $(t,r,\varphi)$ of  \eqref{eq:metric}, $G^{R}$, from the one in the $(T, z, X)$ coordinates, $\tilde{G}^{R}$ in \eqref{eq:Gtilde}, as  
\begin{equation} \label{eq:relation of GR and GRt}
    G^R (\omega, k) = (2\pi T_L)^{\nu-\frac{1}{2}} \,(2\pi T_R)^{\nu+\frac{1}{2}} \,\tilde{G}^R(k_T, k_X)\,,
\end{equation}
where, as we will show shortly, the prefactor $(2\pi T_L)^{\nu-\frac{1}{2}}$ plays an important role to study the pole-skipping in the extreme limit.
Also, comparing $\psi \sim e^{-i \omega t \,+\, i k \varphi}$ with \eqref{eq:fermionic field ansatz}, one can find the relation between $(\omega,k)$ and $(k_T, k_X)$ as
\begin{equation}\label{eq:kTkX}
    k_T + k_X = \frac{\omega + k}{2\pi T_R}\,, \quad k_T - k_X = \frac{\omega - k}{2\pi T_L}\,,
\end{equation}
where we use \eqref{eq:TLTR OmegaT} together with \eqref{eq:TorhoTX}.

In summary, once $\tilde{G}^{R}(k_T, k_X)$ is evaluated via \eqref{eq:Gtilde}, one can obtain $G^{R}(\omega, k)$ by the relation \eqref{eq:relation of GR and GRt}-\eqref{eq:kTkX}. Following this procedure, we present the analytic result of $G^{R}(\omega, k)$ for both the non half-integer case and half integer case below.

\subsubsection{Non half-integer $\nu$ case} \label{subsec:Non half-integer GR Fermi}
Plugging analytic solutions \eqref{eq:chiSol} into \eqref{eq:psiTochi}, one can find the analytic expression for the spinor $\psi_{\pm}$. Furthermore, expanding $\psi_{\pm}$ near the AdS boundary ($z\rightarrow 1$), we can read the coefficients $\mathcal{A}$ and $\mathcal{D}$ as
\begin{align}\label{FRAD}
\begin{split}
    \mathcal{A} & =\sqrt{2}\pi \frac{ \Gamma(a+b-\nu+\frac{1}{2}) \csc{\left(\nu\pi\right)}}{\Gamma(a)\Gamma(b+1)\Gamma( \frac{1}{2}-\nu)}\,,\\
    \mathcal{D} & =-\sqrt{2}\pi \frac{ \Gamma(a+b-\nu+\frac{1}{2}) \csc{\left( \nu\pi \right)}}{\Gamma(a-\nu+\frac{1}{2}) \Gamma(b-\nu+\frac{1}{2}) \Gamma(\frac{1}{2}+\nu) }\,,
\end{split}
\end{align}
where ($a,b$) is given in \eqref{abcre}.
Then, evaluating their ratio, \eqref{eq:Gtilde}, $\tilde{G}^R$ can be simply found as
\begin{align}\label{eq:Gt Non-integer}
\begin{split}
    \tilde{G}^R(k_T, k_X) = -i \frac{\Gamma(\frac{1}{2}-\nu)}{\Gamma(\frac{1}{2}+\nu)}
    \frac{\Gamma({a}_L+\nu-\frac{1}{2})\Gamma({b}_R+\nu+\frac{1}{2})}{\Gamma({a}_L)\Gamma({b}_R)} \,.
\end{split}
\end{align}
Here, we also define
\begin{align}\label{tildeab}
\begin{split}
    a_L & \,:=\, a-\nu+\frac{1}{2} \,=\, \frac{1}{2}\left(\frac{3}{2} - \nu \right) + \frac{\omega - k}{i\,4\pi T_L} \,,  \\
    b_R & \,:=\, b-\nu+\frac{1}{2} \,=\, \frac{1}{2}\left(\frac{1}{2} - \nu \right) + \frac{\omega + k}{i\,4\pi T_R} \,,
\end{split}
\end{align}
where we used \eqref{abcre} together with \eqref{eq:kTkX} in order to express $\tilde{G}^R$ in terms of ($\omega, k$).
Note that the temperature dependence ($T_L, T_R$) of Green's function is encoded in the refined parameters ($a_L, b_R$) in \eqref{tildeab}. 

Therefore, using \eqref{eq:relation of GR and GRt}, we can find the expression of $G^R(\omega,k)$ as
\begin{equation} \label{eq:Fermionic GR Non-Int Finite T}
    G^R(\omega,k) = - i (2\pi T_L)^{\nu-\frac{1}{2}} (2\pi T_R)^{\nu + \frac{1}{2}} \frac{\Gamma(\frac{1}{2}-\nu)}{\Gamma(\frac{1}{2}+\nu)} \frac{\Gamma({a}_L+\nu-\frac{1}{2})}{\Gamma({a}_L)} \frac{\Gamma({b}_R+\nu+\frac{1}{2})}{\Gamma({b}_R)}\,.
\end{equation}
Note that this Green's function is for the non-extremal black hole case.\\

For the case of an extreme limit ($T_L\rightarrow0$), one can notice that $a_L\rightarrow\infty$ in \eqref{tildeab}.
Therefore, in order to describe the extreme limit of $G^R(\omega,k)$, first it is useful to consider the following asympotics of the gamma function 
\begin{equation}\label{ASYMGA}
    \Gamma(x) \,\ra\, \sqrt{2\pi} \, e^{-x} \, x^{x-1/2}\,. \qquad (x \ra \infty)
\end{equation}
Then, using \eqref{ASYMGA}, one can easily check that the ratio between $\Gamma({a}_L+\nu-\frac{1}{2})$ and $\Gamma({a}_L)$ in \eqref{eq:Fermionic GR Non-Int Finite T} can be expressed in $a_L\rightarrow\infty$ limit as
\begin{equation} \label{eq:Gamma function fraction expansion}
\begin{aligned}
    \frac{\Gamma({a}_L+\nu-\frac{1}{2})}{\Gamma({a}_L)}
     \,\approx\, {a}_L^{\nu-\frac{1}{2}} \,\approx\, \left(\frac{\omega - k}{i\, 4\pi T_L}\right)^{\nu-\frac{1}{2}} \,,
\end{aligned}
\end{equation}
where we also used \eqref{tildeab} in the last equality.

Therefore, now we can find the Green's function in the extreme limit, $G^{R}_{0}(\omega,k):=G^R(\omega,k)|_{T_L=0}$, as
\begin{equation} \label{eq:Fermionic GR Non-Int Zero T}
    G^R_0(\omega,k)  =-i (2\pi T_R)^{\nu+\frac{1}{2}} \frac{\Gamma(\frac{1}{2}-\nu)}{\Gamma(\frac{1}{2}+\nu)} \frac{\Gamma({b}_R+\nu+\frac{1}{2})}{\Gamma({b}_R)}\left( \frac{\omega-k}{2i} \right)^{\nu-\frac{1}{2}}\,,
\end{equation}
where the overall prefactor in \eqref{eq:Fermionic GR Non-Int Finite T}, $(2\pi T_L)^{\nu-\frac{1}{2}}$, is canceled out with the one from \eqref{eq:Gamma function fraction expansion}.

\subsubsection{Half-integer $\nu$ case} \label{subsec:Half-integer GR Fermi}

Next we discuss the half-integer case.
For this purpose, it is useful to consider the following property of the hypergeometric function, which can be used in the analytic spinor \eqref{eq:chiSol}.

It is shown \cite{Ceplak:2019ymw} that when 
$n \in \mathbb{Z}^{+}$
the hypergeometric function $_{2}F_{1}(\tilde{a}, \tilde{b}; \tilde{a} + \tilde{b} - n; z)$ can be expressed as
\begin{align}\label{eq:HypergeoPosn}
\begin{split}
&_{2}F_{1}(\tilde{a}, \tilde{b}; \tilde{a} + \tilde{b} - n; z) = \frac{(n - 1)!\Gamma(\tilde{a} + \tilde{b} - n)}{\Gamma(\tilde{a})\Gamma(\tilde{b})}(1-z)^{-n} \sum_{j=0}^{n-1}\frac{(\tilde{a} - n)_{j} (\tilde{b} - n)_{j}(1 - z)^{j}}{j!(1 - n)_{j}} \\
& \quad + (-1)^n\frac{\Gamma(\tilde{a} + \tilde{b} - n)}{\Gamma(\tilde{a} - n)\Gamma(\tilde{b} - n)}  \sum_{j=0}^{\infty} \frac{(\tilde{a})_{j}(\tilde{b})_{j}}{j!(j + n)!}\biggr[-\log(1 - z) + \Psi(j + 1) + \Psi(j + n + 1) \\
& \quad - \Psi(\tilde{a} + j) - \Psi(\tilde{b} + j)\biggr](1 - z)^{j}\,,
\end{split}
\end{align}
where $(\tilde{a})_j:={\Gamma(\tilde{a}+j)}/{\Gamma(\tilde{a})}$ is the Pochhammer symbol and $\Psi(\tilde{a})$ the digamma function. 
Note that hypergeometric functions in \eqref{eq:chiSol} can be expressed by \eqref{eq:HypergeoPosn} for positive integer $n_0=\nu-\frac{1}{2}$ because $\chi_1$ and $\chi_2$ can be written as
\begin{equation}
\begin{aligned}
    \chi_1 \, & \approx \, {}_2 F_{1}(a, b+1; a+(b+1)-n_0; z)\,, \\
    \chi_2 \, & \approx \, {}_2 F_{1}(a, b; a+b-n_0; z)\,.
\end{aligned}
\end{equation}

Therefore, one can expand the analytic spinior $\psi_{\pm}$ near the AdS boundary using \eqref{eq:HypergeoPosn} and find the coefficients in \eqref{eq:Integer boundary behavior} for the half-integer case as
\begin{equation}\label{FRAD2}
\begin{aligned}
    \bar{\mathcal{A}} & = \sqrt{2}\frac{\Gamma(a+b-\nu+\frac{1}{2}) \Gamma(\frac{1}{2}+\nu)}{\Gamma(a) \Gamma(b+1)}\,, \\
    \bar{\mathcal{D}} & = \frac{(-1)^{\nu+1} i}{\sqrt{2}} \frac{\Gamma(a + b -\nu +\frac{1}{2})}{\Gamma(a-\nu +\frac{1}{2}) \Gamma( b -\nu + \frac{1}{2}) \Gamma(\frac{1}{2}+\nu) }\\
    & \hspace{2.5cm} \times \bigg[ 2\Psi(a) + \Psi(b+1) + \Psi(b) - 2\Psi\left(\nu+\frac{1}{2}\right) - 2\Psi(1) \bigg]\,.
\end{aligned}
\end{equation}
Then the Green's function is obtained via \eqref{eq:relation of GR and GRt} together with \eqref{eq:Gtilde} as
\begin{equation} \label{eq:Fermionic GR Int Finite T}
\begin{split}
     {G}^R(\omega, k) =\, & (-1)^\nu (2\pi T_L)^{\nu-\frac{1}{2}} (2\pi T_R)^{\nu+\frac{1}{2}} \frac{1}{2\Gamma(\frac{1}{2}+\nu)^2} \frac{\Gamma({a}_L+\nu-\frac{1}{2})}{\Gamma({a}_L)}\frac{\Gamma({b}_R+\nu+\frac{1}{2})}{\Gamma({b}_R)}\\
     & \times \left[2 \Psi\left({a}_L+\nu-\frac{1}{2}\right) +\Psi\left({b}_R+\nu+\frac{1}{2}\right) +\Psi\left({b}_R+\nu-\frac{1}{2}\right) \right]\,,
\end{split}
\end{equation}
where we use \eqref{tildeab} and ignore the contact terms, $2\Psi\left(\nu+\frac{1}{2}\right)+2\Psi(1)$, not related with pole-skipping.\\

In the extreme limit ($T_L \ra 0$ or $a_L \rightarrow \infty$), we can also use \eqref{eq:Gamma function fraction expansion} even for the half-integer case.
Furthermore, we also make use of the property of the digamma function 
\begin{equation}\label{ASYMDIGA}
    \Psi(x) \ra \log{x}\,, \qquad (x \ra \infty) \,,
\end{equation}
i.e., the digamma function in \eqref{eq:Fermionic GR Int Finite T}, $\Psi\left({a}_L+\nu-\frac{1}{2}\right)$, can be expressed as
\begin{equation}\label{eq:Digamma function expansion}
\begin{aligned}
    \Psi\left({a}_L+\nu-\frac{1}{2}\right)
     \approx \log {a}_L
     \approx \log\frac{\omega-k}{i} \,, 
\end{aligned}
\end{equation}
where we omit a $\log T_L$ term irrelevant for the structure of the pole-skipping.\footnote{Such a log-term also appears in the scalar field analysis~\cite{Natsuume:2020snz} as well as the vector field analysis in the next section. However, as demonstrated in \cite{Natsuume:2020snz}, the absence of the log-term may also be justified from the direct calculation of solving the equations of motion in the extremal background.}
Therefore, the Green's function in the extreme limit becomes
\begin{equation} \label{eq:Fermionic GR Int Zero T}
\begin{split}
    G^R_{0} (\omega, k) =\, & (-1)^\nu (2\pi T_R)^{\nu+\frac{1}{2}} \frac{1}{2\Gamma(\nu+\frac{1}{2})^2} \frac{\Gamma({b}_R+\nu+\frac{1}{2})}{\Gamma({b}_R)} \left( \frac{\omega-k}{2 i} \right)^{\nu-\frac{1}{2}} \\
     & \times \left[ 2 \log\frac{\omega-k}{i} +\Psi\left({b}_R+\nu+\frac{1}{2}\right) +\Psi\left({b}_R+\nu-\frac{1}{2}\right) \right]\,.
\end{split}
\end{equation}

%
\subsection{Fermionic pole-skipping points with rotations}

In this section, we investigate the pole-skipping of the fermionic Green's function: \eqref{eq:Fermionic GR Non-Int Finite T}, \eqref{eq:Fermionic GR Non-Int Zero T} for the non half-integer case and \eqref{eq:Fermionic GR Int Finite T}, \eqref{eq:Fermionic GR Int Zero T} for the half-integer case.

\subsubsection{Non half-integer $\nu$ case} \label{subsec:Non half-integer nu case Fermi}

\paragraph{Pole-skipping in the non-extremal case.}
Let us start with the non half-integer case.
In particular, we first discuss the case of the non-extremal Green's function \eqref{eq:Fermionic GR Non-Int Finite T}.
One can notice that we have two types of poles and zeros from \eqref{eq:Fermionic GR Non-Int Finite T} as:
\begin{align}\label{FPF1}
\begin{split}
    \text{(left poles):} \qquad & {a}_L+\nu-\frac{1}{2} \,=\, \frac{1}{2}\left( \frac{1}{2} + \nu \right) + \frac{\omega-k}{i\,4 \pi T_L} \,=\, - n^p_L\,,\\
    \text{(right poles):} \qquad & {b}_R+\nu+\frac{1}{2} \,=\, \frac{1}{2}\left( \frac{3}{2} +\nu \right) + \frac{\omega+k}{i\, 4 \pi T_R} \,=\, - n^p_R\,,
\end{split}
\end{align}
where $n^p_L,n^p_R=0,1,2,\cdots$ and
\begin{align}\label{FPF2}
\begin{split}
    \text{(left zeros):} \qquad & {a}_L \,=\, \frac{1}{2}\left( \frac{3}{2} - \nu \right)  + \frac{\omega-k}{i\, 4 \pi T_L} \,=\, - n^z_L\,,\\
    \text{(right zeros):} \qquad & {b}_R \,=\, \frac{1}{2}\left(\frac{1}{2} - \nu \right) + \frac{\omega+k}{i\, 4 \pi T_R} \,=\, - n^z_R\,,
\end{split}
\end{align}
where $n^z_L,n^z_R=0,1,2,\cdots$. 

Then, the pole-skipping point at which the pole line intersects with the zero line can be obtained from the following combinations
\begin{align}\label{eq:Fermionic p-s Non-Int Finite T}
\begin{split}
\text{(left poles \& right zeros):}\qquad\qquad\qquad\qquad\qquad \\
i \omega = 2 \pi T_R \left\{ \frac{1}{2} \left(\frac{1}{2} - \nu \right)+n^z_R \right\} &+ 2 \pi T_L \left\{ \frac{1}{2} \left(\frac{1}{2} + \nu \right)+n^p_L \right\}\,,\\
i k = 2 \pi T_R \left\{ \frac{1}{2} \left(\frac{1}{2} - \nu \right)+n^z_R \right\} &- 2 \pi T_L \left\{ \frac{1}{2} \left(\frac{1}{2} + \nu \right)+n^p_L \right\}\,,\\
\text{(right poles \& left zeros):} \qquad\qquad\qquad\qquad\qquad \\
i \omega = 2 \pi T_R \left\{ \frac{1}{2} \left(\frac{3}{2} + \nu \right)+n^p_R \right\} &+ 2 \pi T_L \left\{ \frac{1}{2} \left(\frac{3}{2} - \nu \right)+n^z_L \right\}\,,\\
i k = 2 \pi T_R \left\{ \frac{1}{2} \left(\frac{3}{2} + \nu \right)+n^p_R \right\} &- 2 \pi T_L \left\{ \frac{1}{2} \left(\frac{3}{2} - \nu \right)+n^z_L \right\}\,.
\end{split}
\end{align}
Note that the remaining combination, \text{(left poles \& left zeros)} and \text{(right poles \& right zeros)}, cannot produce the pole-skipping point.\\

It may be instructive to show the leading pole-skipping point to discuss the effect of rotation.
One can check that the first combination, \eqref{eq:Fermionic p-s Non-Int Finite T}, produces the leading pole-skipping point $(\omega_\text{leading}, k_\text{leading})$, in particular using \eqref{eq:TLTR OmegaT}, we find 
\begin{align}\label{LPSPHERM11}
\begin{split}
(\omega_\text{leading}\,,\, k_\text{leading}) = -\frac{2\pi i \,T_h}{1-\Omega^2}\bigg( \frac{1}{2} - \nu \, \Omega \,,\,\,  \frac{\Omega}{2} - \nu \bigg)\,,
\end{split}
\end{align}
where the correction of rotation $\Omega$ appears both in frequency and wave-vector.

\paragraph{Pole-skipping in the extreme limit.}
We also discuss the pole-skipping in the extreme limit ($T_L \rightarrow 0$) for the non half-integer case. Note that such a pole-skipping cannot be achieved simply by taking $T_L \rightarrow 0$ on \eqref{eq:Fermionic p-s Non-Int Finite T} because \eqref{FPF1}-\eqref{FPF2} are ill-defined in the extreme limit.

Instead, we need to consider \eqref{eq:Fermionic GR Non-Int Zero T} in order to discuss the pole-skipping in the extreme limit. 
From the structure of \eqref{eq:Fermionic GR Non-Int Zero T}, one can find the two types of the `would-be' pole-skipping points, ($\omega_{(I)}, k_{(I)}$) and ($\omega_{(II)}, k_{(II)}$), as
\begin{align}\label{WBP1}
\begin{split}
\nu > \frac{1}{2}&:\,\,  \text{right poles} \,\,\,\&\,\,\, (\omega-k)^{\nu - \frac{1}{2}} = 0 \quad\rightarrow\quad \omega_{(I)} = k_{(I)} = -\frac{i \pi T_R}{2} (3 + 4 n^{p}_{R} + 2\nu)\,,\\
\nu < \frac{1}{2}&:\,\,  \text{right zeros} \,\,\,\&\,\,\, (\omega-k)^{\nu - \frac{1}{2}} = \infty \,\,\,\,\rightarrow\quad \omega_{(II)} = k_{(II)} = -\frac{i \pi T_R}{2} (1 + 4 n^{z}_{R}-2\nu)\,,
\end{split} 
\end{align}
where right poles are from \eqref{FPF1} and right zeros from \eqref{FPF2}.

However, as in the scalar field case~\cite{Natsuume:2020snz}, \eqref{WBP1} may not be considered as the pole-skipping points.
Recall that the pole-skipping phenomena indicates that the Green's function near the pole-skipping point cannot give a uniquely-determined finite value due to the slope $\delta \omega/\delta k$ \eqref{VFEQIDP}.
One can check that \eqref{WBP1} does not correspond to this case. For instance, the Green's function \eqref{eq:Fermionic GR Non-Int Zero T} near the `would-be' pole-skipping points \eqref{WBP1} can be expressed as
\begin{align}\label{CHECKINGPSW}
\begin{split}
G^R_0(\omega_{(I)} + \delta \omega, k_{(I)} + \delta k)  &\,\,\approx\,\, \mathcal{F}\left(\frac{\delta \omega}{\delta k}, \nu\right) \delta \omega^{\nu-\frac{3}{2}} \quad\rightarrow\quad
\begin{cases}
       \,\,\, 0 \qquad\,\, \left(  \nu > \frac{3}{2} \right) \,, \\
       \,\, \infty \qquad \left( \frac{1}{2}<\nu<\frac{3}{2} \right) \,,
\end{cases}
\\
G^R_0(\omega_{(II)} + \delta \omega, k_{(II)} + \delta k)  &\,\,\approx\,\, \mathcal{F}\left(\frac{\delta \omega}{\delta k}, \nu\right) \delta \omega^{\nu-\frac{3}{2}} \quad\rightarrow\quad
\begin{cases}
        \,\,\, \infty \qquad \left(0 < \nu<\frac{1}{2} \right) \,.
\end{cases}
\end{split} 
\end{align}
Therefore, \eqref{WBP1} cannot be taken as the pole-skipping points since the Green's function can be solely determined as $\infty$ or $0$.

In other words, one cannot find the pole-skipping points in the extreme limit for the non-half integer case. As we will see shortly, this is not the case for the half-integer case (e.g, $\nu=3/2$).

\subsubsection{Half-integer $\nu$ case and pole-skipping in extreme limit} \label{subsec:Half-integer p-s Fermi}

\paragraph{Pole-skipping in the non-extremal case.} 
For the case of the half-integer case, it is useful to consider the following property of gamma function
\begin{equation}\label{eq:Gamma function ratio}
    \frac{\Gamma(\alpha+n)}{\Gamma(\alpha)}=(\alpha+n-1)\times\cdots \times \alpha \,, \qquad n \in \mathbb{Z}^{+} \,.
\end{equation}

Using \eqref{eq:Gamma function ratio}, one can notice that the gamma functions in \eqref{eq:Fermionic GR Int Finite T} can be expressed as
\begin{align}\label{GMFR12}
\begin{split}
&\frac{\Gamma(a_{L}+\nu-\frac{1}{2})}{\Gamma(a_{L})}=\left(a_{L}+\nu-\frac{3}{2}\right)\times \left(a_{L}+\nu-\frac{5}{2}\right)\times \cdots \times a_{L} \,, \qquad \nu-\frac{1}{2} \in \mathbb{Z}^{+} \,, \\
&\frac{\Gamma(b_{R}+\nu+\frac{1}{2})}{\Gamma(b_{R})}= \left(b_{R}+\nu-\frac{1}{2}\right)\times \left(b_{R}+\nu-\frac{3}{2}\right) \times \cdots \times b_{R} \,, \qquad \nu+\frac{1}{2} \in \mathbb{Z}^{+} \,,
\end{split}
\end{align}
which give the following zeros
\begin{align}\label{zerosnhi1}
\begin{split}
    \text{(left zeros):} \qquad & {a}_L = \frac{1}{2}\left( \frac{3}{2} - \nu \right) +\frac{\omega-k}{i \, 4 \pi T_L} = - n^z_L  \,,\\
    \text{(right zeros):} \qquad & {b}_R = \frac{1}{2}\left(\frac{1}{2} - \nu \right) + \frac{\omega+k}{i \, 4 \pi T_R} = - n^z_R\,,
\end{split}
\end{align}
where $n^z_L=0,1,\cdots, \nu-{3}/{2}$ and $n^z_R=0,1,\cdots, \nu-{1}/{2}$.

One can notice two things for the half-integer case.
First, for the half-integer case, the gamma functions in the Green's function only produce zeros unlike the non half-integer case \eqref{FPF1}-\eqref{FPF2}.
Second, the condition for zeros is restricted by the value of $\nu$: for instance, $n^z_L$ is up to $\nu-3/2$.

On the other hand, poles of \eqref{eq:Fermionic GR Int Finite T} are associated with the digamma functions therein. For our case, one can find poles from the three digamma functions, respectively, as
\begin{align}\label{polesnhi1}
\begin{split}
    \text{(left poles):} \qquad & {a}_L+\nu-\frac{1}{2} = \frac{1}{2}\left(  \frac{1}{2} + \nu\right) + \frac{\omega-k}{i\, 4 \pi T_L} = - n^p_L\,,\\
    \text{(1st right poles):} \qquad & {b}_R+\nu+\frac{1}{2} = \frac{1}{2}\left( \frac{3}{2} + \nu \right) + \frac{\omega+k}{i\, 4 \pi T_R} = - n^p_R\,, \\ 
    \text{(2nd right poles):} \qquad & {b}_R+\nu-\frac{1}{2} = 0  \qquad\&\qquad  \text{(1st right poles)} \,,
\end{split}
\end{align}
where $n^p_L\,,n^p_R = 0,1,2,\cdots$.\\

Before continuing with our analysis, let us pause and discuss the 2nd right poles.
As can be seen, the 2nd right poles have the same structure with the 1st right poles except for the leading condition: $b_R + \nu - \frac{1}{2}=0$.

However, we also find that this leading (``pole") condition may not only be irrelevant for our pole-skipping analysis, but also affect on the range for the zeros: $n_R^z$.
For instance, one can find the same condition from the right ``zeros" when $n_R^z = \nu-1/2$ in \eqref{zerosnhi1}.
This implies that the Green's function \eqref{eq:Fermionic GR Int Finite T} can be expressed as
\begin{equation} \label{DSRBE33}
{G}^R(\omega, k)  \,\approx\, \left(b_{R}+\nu-\frac{1}{2}\right) \Psi\left({b}_R+\nu-\frac{1}{2}\right) \,\approx\, -1 \,,
\end{equation}
when $b_R + \nu - \frac{1}{2}=0$, which is a finite constant. In other words, the condition satisfying $b_R + \nu - \frac{1}{2}=0$ cannot be considered as zeros (or poles). Thus, we need to refine the range for the right zeros as $n^z_R=0,1,\cdots, \nu-{3}/{2}$.\\

In summary, for the half-integer case, we have the (left/right) zeros \eqref{zerosnhi1} and the (left/1st right) poles \eqref{polesnhi1} with 
\begin{align}\label{HICRANGE}
n^z_L \,, n^z_R = 0,1,\cdots, \nu-\frac{3}{2} \,,  \qquad
n^p_L \,, n^p_R = 0,1,2,\cdots \,.
\end{align}
Then, based on zeros and poles obtained above, we can discuss the pole-skipping for the half-integer case.
We find the pole-skipping points as
\begin{align}\label{eq:Fermionic p-s Int Finite T222}
\begin{split}
    \text{(right zeros \& left poles):}
\qquad\qquad\qquad\qquad\qquad \\
        i \omega = 2 \pi T_R \left\{ \frac{1}{2} \left(\frac{1}{2} - \nu \right)+n^z_R \right\} &+ 2 \pi T_L \left\{ \frac{1}{2} \left(\frac{1}{2} + \nu \right)+n^p_L \right\}\,,\\
        i k = 2 \pi T_R \left\{ \frac{1}{2} \left(\frac{1}{2} - \nu \right)+n^z_R \right\} &- 2 \pi T_L \left\{ \frac{1}{2} \left(\frac{1}{2} + \nu \right)+n^p_L \right\}\,, \\
\text{(left zeros \& 1st right poles):}
\qquad\qquad\qquad\qquad\qquad \\
        i \omega = 2 \pi T_R \left\{ \frac{1}{2} \left(\frac{3}{2} + \nu \right)+n^p_R \right\} &+ 2 \pi T_L \left\{ \frac{1}{2} \left(\frac{3}{2} - \nu \right)+n^z_L \right\}\,,\\
        i k = 2 \pi T_R \left\{ \frac{1}{2} \left(\frac{3}{2} + \nu \right)+n^p_R \right\} &- 2 \pi T_L \left\{ \frac{1}{2} \left(\frac{3}{2} - \nu \right)+n^z_L \right\}\,,
\end{split}
\end{align}
which have the same structure with the non half-integer case \eqref{eq:Fermionic p-s Non-Int Finite T}: however, recall that the range of zeros is restricted for the half-integer case \eqref{HICRANGE}. 
Also, notice that the functional form of the leading pole-skipping point is the same with \eqref{LPSPHERM11}.\\

\paragraph{Pole-skipping in the extreme limit.}
Next, let us discuss the extreme limit of the pole-skipping for the half-integer case.
In order for this, we need to consider \eqref{eq:Fermionic GR Int Zero T}
\begin{equation}\label{G0FORDD}
\begin{split}
    G^R_{0} (\omega, k)  \approx \,\, & \frac{\Gamma({b}_R+\nu+\frac{1}{2})}{\Gamma({b}_R)} \left( \omega-k \right)^{\nu-\frac{1}{2}} \\
     & \times \left[ 2 \log (\omega-k) +\Psi\left({b}_R+\nu+\frac{1}{2}\right) +\Psi\left({b}_R+\nu-\frac{1}{2}\right) \right]\,.
\end{split}
\end{equation}
First, we can find that the factor $\frac{\Gamma({b}_R+\nu+\frac{1}{2})}{\Gamma({b}_R)} \left( \omega-k \right)^{\nu-\frac{1}{2}}$ only give the zeros as
\begin{align}
\begin{split}
    \text{(right zeros):} \qquad & {b}_R = \frac{1}{2}\left(\frac{1}{2} - \nu \right) +\frac{\omega+k}{ i\, 4 \pi T_R} = - n^z_R\,, \\
    \text{(left zeros):} \qquad & \omega - k = 0\,,
\end{split}
\end{align}
where \eqref{HICRANGE}.
Note that unlike the non half-integer case \eqref{WBP1}, 
$\left( \omega-k \right)^{\nu-\frac{1}{2}}$ only produces zeros, (left zeros), for the half-integer case since \eqref{nurange2}.\footnote{One may also find the zeros from the rest, $\left[ 2 \log (\omega-k) +\Psi\left({b}_R+\nu+\frac{1}{2}\right) +\Psi\left({b}_R+\nu-\frac{1}{2}\right) \right]$, however we do not consider this case in this paper since such zeros cannot make the pole-skipping phenomena: recall that the factor $\frac{\Gamma({b}_R+\nu+\frac{1}{2})}{\Gamma({b}_R)} \left( \frac{\omega-k}{2 i} \right)^{\nu-\frac{1}{2}}$ cannot produce the pole structure.}

On the other hand, we can find the poles from the rest, $2 \log (\omega-k) +\Psi\left({b}_R+\nu+\frac{1}{2}\right) +\Psi\left({b}_R+\nu-\frac{1}{2}\right)$, as
\begin{align}\label{PEL}
\begin{split}
    \text{(right poles):} \qquad & {b}_R+\nu+\frac{1}{2} = \frac{1}{2}\left( \frac{3}{2} + \nu \right) + \frac{\omega+k}{i\, 4 \pi T_R} = - n^p_R\,.
\end{split}
\end{align}
where \eqref{HICRANGE}.
Notice that the two digamma functions give this right poles for the same reason described around \eqref{DSRBE33}.
Also note that one may also find the poles from $\log (\omega-k)$, $\omega-k=0$, however this cannot be considered as the poles since the structure of \eqref{G0FORDD} shows that $\omega-k=0$ corresponds to the zeros: 
\begin{equation}\label{}
\left( \omega-k \right)^{\nu-\frac{1}{2}}\log (\omega-k) \approx 0  \,,
\end{equation}
when $\omega \rightarrow k$.\\

Finally, combining the left zeros, $\omega-k=0$, with the right poles \eqref{PEL}, we find the pole-skipping point ($\omega_{0}, k_{0}$) as
\begin{align}\label{PSPHIEXT}
\omega_{(0)} = k_{(0)} = -\frac{i \pi T_R}{2} (3 + 4 n^{p}_{R} + 2\nu) \,.
\end{align}
Furthermore, as did in the non half-integer case \eqref{CHECKINGPSW}, we can expand the Green's function \eqref{eq:Fermionic GR Int Zero T} near \eqref{PSPHIEXT} and find
\begin{align}
G^R_0(\omega_{(0)} + \delta \omega, k_{(0)} + \delta k)  &\quad\approx\quad \mathcal{F}
\left(\frac{\delta \omega}{\delta k}, \nu\right)
\delta \omega^{\nu-\frac{3}{2}}  \,,
\end{align}
which implies that \eqref{PSPHIEXT} can be considered as the pole-skipping point when $\nu=3/2$.
Therefore, we can find the pole-skipping point even in the extreme limit for the half-integer case when $\nu=3/2$ as:
\begin{align}\label{answer111}
\left(\omega^{\text{ext}}\,,\, k^{\text{ext}}\right):= \left( \omega_{(0)}, k_{(0)} \right) \Big|_{\nu=3/2} = -2 \pi i T_R \left(\frac{3}{2}+n^p_R\right) \,,
\end{align}
in which its leading is 
\begin{align}\label{}
\left( \omega^{\text{ext}}_\text{leading}\,,\, k^{\text{ext}}_\text{leading} \right) = -3 \pi i T_R \,.
\end{align}

%
\section{Vector fields ($s=1$)} \label{sec:Vector fields}
In this section, our focus is on the pole-skipping phenomena of massive vector fields and their response to rotations.
We begin by employing the approach used in computing the analytic solution for vector fields \cite{Das:1999pt, Birmingham:2001pj}, enabling us to derive the analytic Green's function and investigate the pole-skipping behavior. Although the pole-skipping points for vector fields have been previously studied, they mostly have been explored in the context of non-rotating black holes, for instance \cite{Natsuume:2019xcy}. To the best of our knowledge, the literature lacks both the explicit form of analytic Green's function of the massive vector field and an examination of pole-skipping thereof in the rotating BTZ black hole.

\subsection{Maxwell equations of motion}
The analytic solution of the vector field $A$ for the massive Maxwell equation is given in \cite{Das:1999pt, Birmingham:2001pj}. Based on these papers, we review the analytic solutions for the gauge field in rotating BTZ. The field $A$ satisfies
\begin{equation} \label{eq:Gauge eqn}
    \nabla^\nu\partial_{[\nu}A_{\lambda]}=m_A^2 A_{\lambda}\,,
\end{equation}
with the mass $m_A$. In the coordinates $(T, \rho, X)$ represented in \eqref{eq:metric TrhoX}, as in the fermionic field \eqref{eq:fermionic field ansatz}, the gauge field fluctuation can be written as 
\begin{equation} \label{eq:Gauge field ansatz}
    A = \Big( A_{T}(\rho) \, \td T + A_{\rho}(\rho) \, \td \rho + A_{X}(\rho) \, \td X \Big) e^{-i k_T T + i k_X X}\,.
\end{equation}
In order to solve the equation \eqref{eq:Gauge eqn} analytically, it is convenient to use the ansatz
\begin{equation} \label{eq:ATX}
    A_{T}=\frac{1}{2}\left(A_1 + A_2\right) \,, \quad A_{X}=\frac{1}{2}\left(A_1 - A_2\right)\,.
\end{equation}
Using this ansatz, the equations of motion in \eqref{eq:Gauge eqn} are decoupled as
\begin{equation}
    \partial^2 A_i + (\tanh{\rho} + \coth{\rho})\partial_{\rho}A_i = (m_A^2-2\epsilon_i m_A)A_i\,,
\end{equation}
for $i=1,2$. Here, $\partial^2=-\text{csch}^2{\rho} \, \partial_{T}^2+\partial_{\rho}^2+\text{sech}^2{\rho}^2 \, \partial_{X}^2$ and $\epsilon_{1,2}=\mp 1$. 
Within the $z$-coordinate introduced in \eqref{eq:psiTochi}, the decoupled equations above can be expressed as
\begin{equation} \label{eq:Decoupled Gauge eqn1}
    z(1-z)\partial_z^{2}A_i + (1-z)\partial_z A_i + \left[ \frac{k_T^2}{4z} - \frac{k_X^2}{4} - \frac{m_A^2-2\epsilon_i m_A}{4(1-z)} \right] A_i = 0\,.
\end{equation}
Furthermore, the remaining equation of motion for $A_\rho$ can be derived from \eqref{eq:Gauge eqn} as
\begin{equation} \label{eq:Arho}
\begin{aligned}
    A_\rho 
    = i\left( \sqrt{z} - \frac{1}{\sqrt{z}} \right) \sum_{j=1,2} \frac{k_X - \epsilon_j k_T}{2 m_A}A_j\,,
\end{aligned}
\end{equation}
where the relation between ($A_T,A_X$) and ($A_1,A_2$) in \eqref{eq:ATX} are used with the definition of $z$.
It should be noted that this relation arises when considering a finite value of $m_A$.
The \textit{analytic} incoming solutions of the equations in \eqref{eq:Decoupled Gauge eqn1} are
\begin{align}\label{eq:A12Sol}
\begin{split}
    A_1 &= e_1 z^{\alpha} (1-z)^{\beta + 1} {}_2 F_{1}(a+1, b+1; c; z)\,, \\
    A_2 &= e_2 z^{\alpha} (1-z)^{\beta} {}_2 F_{1}(a, b; c; z)\,,
\end{split}
\end{align}
where
\begin{align}\label{abcre Gauge}
\begin{split}
    a &=\frac{m_A}{2}-\frac{i}{2}(k_T - k_X)\,, \quad \qquad b =\frac{m_A}{2}-\frac{i}{2}(k_T + k_X)\,, \\
    c &= \frac{1}{2}+a+b-m_A = 1-i k_T\,,
\end{split}
\end{align}
and
\begin{equation}
    \alpha=- \frac{i k_T}{2}, \quad \beta=\frac{m_A}{2}\,.
\end{equation}
These parameters are different with \eqref{abcre} in the fermionic field part. The constants $e_1$ and $e_2$ have a relation
\begin{equation} \label{eq:e1e2 relation}
    \frac{e_2}{e_1} = \frac{i(k_T + k_X) + m_A}{i(k_T - k_X) - m_A} = \frac{b-m_A}{a}\,,
\end{equation}
by Lorenz gauge($\nabla^\nu A_\nu =0$) and the representation theory of one forms on SL(2,R) manifolds \cite{Das:1999pt, Birmingham:2001pj}.\footnote{
The representation theory suggests the equation of motion($\epsilon_\lambda^{\alpha \beta}\partial_\alpha A_\beta = - m_A A_{\lambda}$), because the covering group of AdS spacetime is $\text{SL(2,R)}\times\text{SL(2,R)}$. The rotating BTZ spacetime in \eqref{eq:metric} is locally AdS spacetime.
}


\subsection{Exact vector Green's function} \label{Exact Green's function in spin 1}
\subsubsection{Holographic dictionary for vector Green's function}
Using the analytic solution $A_{1,2}$ obtained above, we study the vector Green's function. Let us first review how to read such a vector boundary correlator.
\paragraph{Non integer case.}
According to the holographic dictionary, the AdS boundary behavior of the gauge field $A$ are also related to the Green's function. For the equations in \eqref{eq:Decoupled Gauge eqn1} within the z-coordinate, one can find the following asymptotic behavior near the AdS boundary $(z\ra1)$
\begin{equation}
\begin{aligned} \label{eq:Gauge Non-Integer boundary behavior 12}
    A_1 &\sim \mathcal{A}(1-z)^{-\frac{1-\Delta_-}{2}} + \mathcal{B} (1-z)^{\frac{1+\Delta_+}{2}} + \cdots\,,\\
    A_2 &\sim \mathcal{C}(1-z)^{\frac{1+\Delta_-}{2}} + \mathcal{D} (1-z)^{-\frac{1-\Delta_+}{2}} + \cdots\,.
\end{aligned}
\end{equation}
where $\Delta_{\pm}$ is the conformal dimension
\begin{equation} \label{SDR Vector}
    \Delta_{\pm} = 1 \pm m_A\,.
\end{equation}

Furthermore, one can also study the asymptotic behavior for the gauge fields in terms of ($t, r, \varphi$) coordinates as follows.
The gauge fields ($A_t,A_r,A_\varphi$) are defined as
\begin{equation} \label{eq:Gauge field ansatz 2}
    A = \Big(A_{t}(r) \td t + A_{r}(r) \td r + A_{\varphi}(r) \td \varphi \Big) e^{-i \omega t + i k \varphi}\,,
\end{equation}
of $(t,r,\varphi)$ coordinates in the metric \eqref{eq:metric}. Comparing with the definition of ($A_T,A_\rho,A_X$) in \eqref{eq:Gauge field ansatz}, below relations are computed.
\begin{equation}\label{eq:Gauge field transform}
    A_t = r_+ A_T - r_- A_X\,, \qquad A_r = A_\rho \left(\frac{\dd \rho}{\dd r}\right) \,, \qquad A_\varphi = r_+ A_X - r_- A_T\,,
\end{equation}
where we use the coordinate transformation \eqref{eq:TorhoTX}. According to the transformations \eqref{eq:ATX}, \eqref{eq:Arho} and \eqref{eq:Gauge field transform}, the near boundary asymptotic behavior for the gauge fields ($A_t,A_r,A_\varphi$) are obtained as
\begin{equation}\label{eq:Gauge Non-Integer boundary behavior}
\begin{split}
    A_t &\sim \mathcal{E} (1-z)^{-\frac{1-\Delta_-}{2}} + \cdots + \mathcal{F} (1-z)^{-\frac{1-\Delta_+}{2}} + \cdots\,,\\
    A_\varphi &\sim \mathcal{E} (1-z)^{-\frac{1-\Delta_-}{2}} + \cdots - \mathcal{F} (1-z)^{-\frac{1-\Delta_+}{2}} + \cdots\,,\\
    A_r &\sim \frac{1}{m_A} \left(\frac{\dd \rho}{\dd r}\right) \left\{\mathcal{H} (1-z)^{\frac{1+\Delta_-}{2}} + \cdots + \mathcal{I} (1-z)^{\frac{1+\Delta_+}{2}} + \cdots\right\}\,,
\end{split}
\end{equation}
where
\begin{equation} \label{eq:HJrelation}
\begin{split}
    \mathcal{H}=-i\,\frac{k_T + k_X}{2\pi T_L}\mathcal{E}\,, \qquad
    &\mathcal{I}=i\,\frac{k_T - k_X}{2\pi T_R}\mathcal{F}\,.
\end{split}
\end{equation}
For the obtained incoming analytic solution ($A_1,A_2$) in \eqref{eq:A12Sol}, the coefficients $\mathcal{E}$ and $\mathcal{F}$ can be found as \eqref{eq:Gauge Solution Behavior Non-Int}. However, the behavior \eqref{eq:Gauge Non-Integer boundary behavior} is not correct for the following integer case.

\paragraph{Integer case.}
As same as \eqref{nudef}, we define the parameter $\nu$ as
\begin{equation}\label{nudef Gauge}
    \nu:=\frac{\Delta_+ - \Delta_-}{2}=m_A\,,
\end{equation}
by the conformal dimension \eqref{SDR Vector}. In particular, using $\nu$, we can rewrite the boundary expansion \eqref{eq:Gauge Non-Integer boundary behavior 12} as
\begin{equation}
\begin{aligned}
    A_1 &\sim \mathcal{A}(1-z)^{-\frac{1-\Delta_-}{2}} + \mathcal{B} (1-z)^{-\frac{1-\Delta_-}{2} + (\nu+1)} + \cdots\,,\\
    A_2 &\sim \mathcal{C}(1-z)^{-\frac{1-\Delta_+}{2}-(\nu-1)} + \mathcal{D} (1-z)^{-\frac{1-\Delta_+}{2}} + \cdots\,.
\end{aligned}
\end{equation}
It is assumed that $\nu$ is not an integer for the boundary expansion \eqref{eq:Gauge Non-Integer boundary behavior 12}:
\begin{equation}
    \nu \pm 1 \,\notin\, \mathbb{Z}^{+}\,.
\end{equation}
In principle, it is also possible to consider the negative integer. However, it is enough to consider the positive integer case for our purpose: see the description around \eqref{nurange2 Vector}.

Therefore, for the case of integer $\nu$, such as
\begin{equation} \label{eq:Int nu range vector}
    \nu=2,~3,~4,~\cdots\,,
\end{equation}
one can find the different boundary behaviors
\begin{equation}\label{eq:Gauge Integer boundary behavior 12}
\begin{aligned}
    A_1 &\sim \bar{\mathcal{A}} (1-z)^{-\frac{1-\Delta_{-}}{2}} \left\{1 + \cdots +\mathcal{O}\left( (1-z)^{\nu+1} \ln{(1-z)}\right)  \right\} + \bar{\mathcal{B}} (1-z)^{\frac{1+\Delta_{+}}{2}} \left\{1+\cdots\right\} \,,
    \\
    A_2 &\sim \bar{\mathcal{C}} (1-z)^{\frac{1+\Delta_{-}}{2}} \left\{1 +\cdots+ \mathcal{O}\left( (1-z)^{\nu-1} \ln{(1-z)}\right) \right\} + \bar{\mathcal{D}} (1-z)^{-\frac{1-\Delta_{+}}{2}} \left\{1+\cdots\right\} \,.
\end{aligned}
\end{equation}
By the transformations \eqref{eq:ATX}, \eqref{eq:Arho} and \eqref{eq:Gauge field transform}, the boundary asymptotic behaviors of ($A_t, A_r, A_\varphi$) are computed from the behaviors of ($A_1,A_2$) in \eqref{eq:Gauge Integer boundary behavior 12} as
\begin{equation} \label{eq:Gauge Integer boundary behavior}
\begin{aligned}
    A_t & \sim \bar{\mathcal{E}}(1-z)^{-\frac{1-\Delta_-}{2}} \left\{1 + \cdots +\mathcal{O}\left( (1-z)^{\nu} \ln{(1-z)}\right)  \right\} + \bar{\mathcal{F}}(1-z)^{-\frac{1-\Delta_+}{2}} \left\{1 + \cdots \right\}\,,\\
    A_\varphi & \sim \bar{\mathcal{E}}(1-z)^{-\frac{1-\Delta_-}{2}} \left\{1 + \cdots +\mathcal{O}\left( (1-z)^{\nu} \ln{(1-z)}\right)  \right\} - \bar{\mathcal{F}}(1-z)^{-\frac{1-\Delta_+}{2}} \left\{1 + \cdots \right\}\,,\\
    A_r & \sim \frac{1}{m_A} \left(\frac{\dd \rho}{\dd r}\right) \bigg\{ \bar{\mathcal{H}}(1-z)^{\frac{1+\Delta_-}{2}} \left\{1 + \cdots +\mathcal{O}\left( (1-z)^{\nu} \ln{(1-z)}\right)  \right\}\\
    & \hspace{8.65cm} + \bar{\mathcal{I}}(1-z)^{\frac{1+\Delta_+}{2}} \left\{1 + \cdots \right\} \bigg\}\,.
\end{aligned}
\end{equation}
We also found the relations corresponding to \eqref{eq:HJrelation}. However, its expression is not so illuminating so we do not display it here.
By using the incoming analytic solution in \eqref{eq:A12Sol}. one can find the analytic expression for $\bar{\mathcal{E}}$, $\bar{\mathcal{F}}$, $\bar{\mathcal{H}}$, and $\bar{\mathcal{J}}$ as \eqref{eq:Gauge Solution Behavior Int} for integer case.

\paragraph{Vector Green's function in holography.} According to the holographic principle, the Green's functions are given by the ratio between two independent parameters as
\begin{equation} \label{eq:Gauge Gtilde}
\begin{aligned}
    \tilde{G}^R_{tt} \,\approx\, \tilde{G}^R_{\varphi \varphi}  \,\approx\, \frac{\mathcal{F}}{\mathcal{E}} \,\,\, \text{or} \,\,\, \frac{\bar{\mathcal{F}}}{\bar{\mathcal{E}}}\,, \qquad
    \tilde{G}^R_{rr}  \,\approx\, \frac{\mathcal{I}}{\mathcal{H}} \,\,\, \text{or} \,\,\, \frac{\bar{\mathcal{I}}}{\bar{\mathcal{H}}}\,,
\end{aligned}
\end{equation}
where the former one is for the non integer case, while the later is for the integer case. Strictly speaking, the Green's functions in \eqref{eq:Gauge Gtilde} corresponds to the case of the standard quantization where $\mathcal{E}$ (or $\bar{\mathcal{E}}$), $\mathcal{H}$ (or $\bar{\mathcal{H}}$) are interpreted as the sources and $\mathcal{F}$ (or $\bar{\mathcal{F}}$), $\mathcal{I}$ (or $\bar{\mathcal{I}}$) are corresponding vevs.

As same as the Fermionic Green's function, it is also possible to consider the other quantization, alternative quantization, by replacing $\mathcal{E}\leftrightarrow \mathcal{F}$ and $\mathcal{H}\leftrightarrow \mathcal{I}$.
However, in this paper, we only consider the case of standard quantization \eqref{eq:Gauge Gtilde} for our own purpose: notice that the structure of the pole-skipping, $\tilde{G}^{R}\sim0/0$, is independent of the type of a quantization.

Equivalently, considering a standard quantization implies that we only focus on the positive $\nu$, for instance, one can notice that the role of $\mathcal{E}$ and $\mathcal{F}$ ($\mathcal{H}$ and $\mathcal{I}$) in \eqref{eq:Gauge Non-Integer boundary behavior} can be exchanged when the $\Delta_{\pm}$ in \eqref{SDR Vector} is defined reversely with the reversed sign of $\nu$ in \eqref{nudef Gauge}.
Therefore, in this paper, we only choose $\nu > 0$ hereafter\footnote{We don't care $\nu=0$ case because $A_{\rho}$ in \eqref{eq:Arho} is ill-defined.}, i.e.,
\begin{align}\label{nurange2 Vector}
\nu = 
\begin{cases}
       \,\, \textrm{Otherwise} \qquad\,\, (\textrm{Non integer case}) \,, \\
       \,\, 2\,,\, 3 \,,\, 4 \,,\, \cdots\,\,\,\, \quad (\textrm{Integer case}) \,.
\end{cases}
\end{align}
Mathematically, it is also possible to consider $\nu=1$. However, this special case may produce ill-defined boundary expansions. In this case, the coefficients $\bar{\mathcal{C}}$ and $\bar{\mathcal{D}}$ of \eqref{eq:Gauge Integer boundary behavior 12} are ill-defined because $\frac{1+\Delta_-}{2}=-\frac{1-\Delta_+}{2}$ for $\nu=1$. As explained around \eqref{nurange2}, we do not consider the special case in this paper.

Furthermore, similar to section \ref{Exact fermionic Green's function}, following the coordinate transformation \eqref{eq:TorhoTX}, one can restore the Green's function in the original coordinates $(t,r,\varphi)$ of \eqref{eq:metric}, $G^R$, from the one in the coordinates $(T,z,X)$, $\tilde{G}^R$ in \eqref{eq:Gauge Gtilde}, as
\begin{equation} \label{eq:relation of GR and GRt Gauge}
    G^R (\omega, k) = (2\pi T_L)^{\nu} \,(2\pi T_R)^{\nu} \,\tilde{G}^R(k_T, k_X)\,.
\end{equation}
The relation between $(\omega, k)$ and $(k_T, k_X)$ is equally given as \eqref{eq:kTkX} by comparing \eqref{eq:Gauge field ansatz} with \eqref{eq:Gauge field ansatz 2}.


\subsubsection{Non integer $\nu$ case}
Plugging the analytic solutions $A_{1,2}$ \eqref{eq:A12Sol} into \eqref{eq:ATX}, one can find the analytic expressions for $A_T$ and $A_X$. Using the solution of $A_\rho$ \eqref{eq:Arho} and transformation \eqref{eq:Gauge field transform}, the analytic solutions for $(A_t, A_r, A_\varphi)$ can be computed. Furthermore, expanding $(A_t, A_r, A_\varphi)$ near the AdS boundary $(z \ra \infty)$, we can read the coefficients in \eqref{eq:Gauge Non-Integer boundary behavior} as
\begin{equation} \label{eq:Gauge Solution Behavior Non-Int}
\begin{aligned}
    \mathcal{E} & = -e_1 \pi^2 T_L \mathcal{K}\,,
    & \mathcal{F} = e_2 \pi^2 T_R \mathcal{L}\,, \hspace{1.4cm}\\
    \mathcal{H} & = -e_1 \pi \left( b-\frac{m_A}{2} \right) \mathcal{K}\,, \qquad
    &  \mathcal{I} = -e_2 \pi \left( a-\frac{m_A}{2} \right) \mathcal{L}\,,
\end{aligned}
\end{equation}
where
\begin{equation}
    \mathcal{K} = \frac{\Gamma(a+b-m_A+1) \csc( m_A \pi)}{\Gamma(a+1) \Gamma(b+1) \Gamma(-m_A)}\,, \qquad \mathcal{L} = \frac{\Gamma(a+b-m_A+1) \csc( m_A \pi)}{\Gamma(a-m_A+1) \Gamma(b-m_A+1) \Gamma(m_A)}\,,
\end{equation}
and $(a,b)$ is given in \eqref{abcre Gauge}. Then, evaluating their ratio, \eqref{eq:Gauge Gtilde}, the analytic Green's functions can be simply found as
\begin{equation}
\begin{aligned}
    \tilde{G}_{tt}^R(k_T, k_X) \, & \approx\, \frac{T_R}{T_L} \frac{\Gamma(-m_A)}{\Gamma(m_A)} \frac{\Gamma(a)\Gamma(b+1)}{\Gamma(a-m_A+1)\Gamma(b-m_A)}\,,\\
    \tilde{G}_{\varphi \varphi}^R(k_T, k_X) \, & \approx \, \tilde{G}_{tt}^R(k_T, k_X)\,,\\
    \tilde{G}_{rr}^R(k_T, k_X) \, & \approx\, \frac{T_L}{T_R}\frac{k_T-k_X}{k_T + k_X} \tilde{G}_{tt}^R(k_T, k_X)\,.
\end{aligned}
\end{equation}
Using the coordinate transformation \eqref{eq:relation of GR and GRt Gauge} and definition of $\nu$ \eqref{nudef Gauge}, we can find the expression of $G^R(\omega,k)$ as
\begin{equation} \label{eq:Gauge GR Non-Int Finite T}
\begin{aligned}
    G^R_{tt}(\omega,k) \, & \approx\, (2\pi T_L)^{\nu-1} (2\pi T_R)^{\nu+1} \frac{\Gamma(-\nu)}{\Gamma(\nu)} \frac{\Gamma(a_L+\nu-1)}{\Gamma(a_L)} \frac{\Gamma(b_R+\nu+1)}{\Gamma(b_R)}\,,\\ G^R_{\varphi \varphi}(\omega,k) \, & \approx \, G^R_{tt}(\omega,k)\,,\\
    G^R_{rr}(\omega,k) \, & \approx\, \frac{\omega-k}{\omega+k} G^R_{tt}(\omega,k)\,.
\end{aligned}
\end{equation}
Here, we define
\begin{equation} \label{tildeab Gauge}
\begin{aligned}
    a_L & := a-\nu+1 \, = \, 1 -\frac{\nu}{2} + \frac{\omega - k}{i\,4\pi T_L}\,,\\
    b_R & := b-\nu = -\,\frac{\nu}{2} + \frac{\omega + k}{i\,4\pi T_R}\,,
\end{aligned}
\end{equation}
where we used together \eqref{abcre Gauge} with \eqref{eq:kTkX} in order to express $G^R$ in terms of $(\omega,k)$. Note that the temperature dependence $(T_L, T_R)$ of Green's function is encoded in the refined parameters $(a_L, b_R)$ in \eqref{tildeab Gauge}. Also, note that the Green's functions \eqref{eq:Gauge GR Non-Int Finite T} are for the non-extremal case.

For the case of an extreme limit, $T_{L} \ra 0$, one can notice that $a_L \ra \infty$ in \eqref{tildeab Gauge}. Similar to section \ref{subsec:Non half-integer GR Fermi}, in order to describe the extreme limit of $G^R(\omega,k)$, it is useful to consider the asymptotic gamma function \eqref{ASYMGA}. Then, one can easily check that the ratio between $\Gamma(a_L + \nu -1)$ and $\Gamma(a_L)$ in \eqref{eq:Gauge GR Non-Int Finite T} can be expressed in $a_L \ra \infty$ limit as
\begin{equation} \label{eq:Gamma function fraction expansion Gauge}
    \frac{\Gamma(a_L + \nu -1)}{\Gamma(a_L)} \, \approx \, a_L^{\nu-1} \, \approx \, \left( \frac{\omega - k}{i \, 4\pi T_L} \right)^{\nu-1}\,,
\end{equation}
where we also used \eqref{tildeab Gauge} in the last equality.

Therefore, now we can find the Green's function in the extreme limit, $G^{R}_{0}(\omega,k):=G^R(\omega,k)|_{T_L=0}$, as
\begin{equation} \label{eq:Gauge GR Non-Int Zero T}
\begin{aligned}
    G^R_{tt,0}(\omega,k) \, & \approx \, (2\pi T_R)^{\nu+1} \frac{\Gamma(-\nu)}{\Gamma(\nu)} \frac{\Gamma(b_R+\nu+1)}{\Gamma(b_R)}\left( \frac{\omega-k}{2i} \right)^{\nu-1}\,,\\
    G^R_{\varphi\varphi,0}(\omega,k) \, & \approx \, G^R_{tt}(\omega,k)|_{T_L=0} \,,\\
    G^R_{rr,0}(\omega,k) \, & \approx \, (2\pi T_R)^{\nu+1} \frac{\Gamma(-\nu)}{\Gamma(\nu)} \frac{\Gamma(b_R+\nu+1)}{\Gamma(b_R)}\left( \frac{\omega-k}{2i} \right)^{\nu} \left( \frac{2i}{\omega+k} \right)\,,
\end{aligned}
\end{equation}
where the overall prefactor in \eqref{eq:Gauge GR Non-Int Finite T}, is canceled out with the one from \eqref{eq:Gamma function fraction expansion Gauge}.

\subsubsection{Integer $\nu$ case}
Next we discuss the integer case. For this purpose, it is useful to consider the property of the hypergeometric function \eqref{eq:HypergeoPosn}, similar to section \ref{subsec:Half-integer GR Fermi}. Note that the hypergeometric functions in \eqref{eq:A12Sol} can be expressed by \eqref{eq:HypergeoPosn} for integer $n_{1,2}$ as
\begin{equation}
    n_{1}=\nu + 1\,, \qquad n_{2}=\nu - 1\,,
\end{equation}
because $A_1$ and $A_2$ can be written as
\begin{equation}
\begin{aligned}
    A_1 \, & \approx \, {}_2 F_{1}(a+1, b+1; (a+1)+(b+1)-n_1; z)\,, \\
    A_2 \, & \approx \, {}_2 F_{1}(a, b; a+b-n_2; z)\,.
\end{aligned}
\end{equation}

Therefore, one can expand the analytic solutions for $(A_t, A_r, A_\varphi)$ near the AdS boundary using \eqref{eq:HypergeoPosn} and find the coefficients in \eqref{eq:Gauge Integer boundary behavior} for the integer case as
\begin{equation} \label{eq:Gauge Solution Behavior Int}
\begin{aligned}
    \bar{\mathcal{E}} & = e_1 \pi T_L \bar{\mathcal{K}}\,,
    & \bar{\mathcal{F}} = (-1)^{m_A} e_2 \pi T_R \bar{\mathcal{L}}\,, \hspace{1.6cm}\\
    \bar{\mathcal{H}} & = e_1 \left( b-\frac{m_A}{2} \right) \bar{\mathcal{K}}\,, \qquad
    & \bar{\mathcal{I}} = (-1)^{m_A+1} e_2 \left( a-\frac{m_A}{2} \right) \bar{\mathcal{L}}\,,
\end{aligned}
\end{equation}
where
\begin{equation}
\begin{aligned}
    \bar{\mathcal{K}} & = \frac{\Gamma(m_A+1)\Gamma(a+b-m_A+1)}{\Gamma(a+1) \Gamma(b+1)}\,, \\
    \bar{\mathcal{L}} & = \frac{\Gamma(a+b-m_A+1)}{\Gamma(m_A)\Gamma(a-m_A+1) \Gamma(b-m_A+1)}\left[ \Psi(a) + \Psi(b) - \Psi(m_A) -\Psi(1) \right] \,.
\end{aligned}
\end{equation}
Then, the Green's functions are obtained via \eqref{eq:relation of GR and GRt Gauge} together with \eqref{eq:Gauge Gtilde} and \eqref{eq:e1e2 relation} as
\begin{equation} \label{eq:Gauge GR Int Finite T}
\begin{aligned}
    G^R_{tt}(\omega, k) \, \approx \, & (2\pi T_L)^{\nu-1} (2\pi T_R)^{\nu+1} \frac{1}{\Gamma(\nu)\Gamma(\nu+1)} \\
    & \times \frac{\Gamma(a_L+\nu-1)}{\Gamma(a_L)} \frac{\Gamma(b_R+\nu+1)}{\Gamma(b_R)} \left[ \Psi(a_L+\nu-1) +\Psi(b_R+\nu) \right]\,,\\
    G^R_{\varphi \varphi}(\omega, k) \,\approx \, & G^R_{tt}(\omega, k)\,, \\
    G^R_{rr}(\omega, k) \,\approx \, & (2\pi T_L)^{\nu-1} (2\pi T_R)^{\nu+1} \frac{1}{\Gamma(\nu)\Gamma(\nu+1)}\frac{\omega-k}{\omega+k} \\
    & \times \frac{\Gamma(a_L+\nu-1)}{\Gamma(a_L)} \frac{\Gamma(b_R+\nu+1)}{\Gamma(b_R)} \left[ \Psi(a_L+\nu-1) +\Psi(b_R+\nu) \right]\,,
\end{aligned}
\end{equation}
where we use \eqref{nudef Gauge} and \eqref{tildeab Gauge} and ignore the contact terms, $\Psi(\nu)+\Psi(1)$, not related with pole-skipping.

In the extreme limit ($T_{L} \ra 0$ or $a_L \ra \infty$), we can use \eqref{eq:Gamma function fraction expansion Gauge} even for the integer case. Similar to section \ref{subsec:Half-integer GR Fermi}, we also use the property of the digamma function \eqref{ASYMDIGA}. Then, the digamma functions in \eqref{eq:Gauge GR Int Finite T} can be expressed as
\begin{equation}
    \Psi(a_L+\nu-1) \, \approx \, \log a_L \, \approx \, \log \frac{\omega-k}{i}\,,
\end{equation}
where we also omit $\log T_L$ irrelevant for structure of the pole-skipping. Therefore, the Green's functions in the extreme limit become
\begin{equation} \label{eq:Gauge GR Int Zero T}
\begin{aligned}
    G^R_{tt,0}(\omega, k) \, \approx \, & (2\pi T_R)^{\nu+1} \frac{1}{\Gamma(\nu)\Gamma(1+\nu)}\\
    & \times \frac{\Gamma(b_R+\nu+1)}{\Gamma(b_R)} \left( \frac{\omega-k}{2i} \right)^{\nu-1} \left[\log{\frac{\omega-k}{i}} +\Psi(b_R+\nu) \right]\,,\\
    G^R_{\varphi \varphi,0}(\omega, k) \,\approx \, & G^R_{tt}(\omega, k)\,, \\
    G^R_{rr,0}(\omega, k) \,\approx \, & (2\pi T_R)^{\nu+1} \frac{1}{\Gamma(\nu)\Gamma(\nu+1)}\\
    & \times \frac{\Gamma(b_R+\nu+1)}{\Gamma(b_R)} \left( \frac{\omega-k}{2i} \right)^{\nu}\left( \frac{2i}{\omega+k} \right) \left[\log{\frac{\omega-k}{i}} + \Psi(b_R+\nu) \right]\,.
\end{aligned}
\end{equation}

\subsection{Pole-skipping points of $G^R_{tt}$ and $G^R_{\varphi \varphi}$}
In this section, we investigate the pole-skipping of the vector Green's function $G^R_{tt}$ and $G^R_{\varphi \varphi}$: \eqref{eq:Gauge GR Non-Int Finite T}, \eqref{eq:Gauge GR Non-Int Zero T} for the non integer case and \eqref{eq:Gauge GR Int Finite T}, \eqref{eq:Gauge GR Int Zero T} for the half-integer case. Note that we focus on $G^R_{tt}$ in this section because $G^R_{\varphi \varphi}$ is obtained by $G^R_{tt}$ for all the cases.

\subsubsection{Non integer $\nu$ case}
\paragraph{Pole-skipping in the non-extremal case.}
We first discuss the case of the non-extremal Green's function $G^R_{tt}$ in \eqref{eq:Gauge GR Non-Int Finite T}. One can notice that we have two types of poles and zeros from \eqref{eq:Gauge GR Non-Int Finite T} as
\begin{align} \label{FPF1 Gauge tt}
\begin{split}
    \text{(left poles):} \qquad & a_L+\nu-1 = \frac{\nu}{2} + \frac{\omega-k}{i\, 4 \pi T_L} = - n^p_L\,,\\
    \text{(right poles):} \qquad & b_R+\nu+1 = 1+\frac{\nu}{2} + \frac{\omega+k}{i\, 4 \pi T_R} = - n^p_R\,,
\end{split}
\end{align}
where $n^p_L,n^p_R=0,1,2,\cdots$ and
\begin{align} \label{FPF2 Gauge tt}
\begin{split}
    \text{(left zeros):} \qquad & a_L = 1-\frac{\nu}{2} + \frac{\omega-k}{i\, 4 \pi T_L} = - n^z_L\,,\\
    \text{(right zeros):} \qquad & b_R = - \frac{\nu}{2} + \frac{\omega+k}{i\, 4 \pi T_R} = - n^z_R\,,
\end{split}
\end{align}
where $n^z_L,n^z_R=0,1,2,\cdots$.

Then, the pole-skipping point at which the pole line intersects with the zero line can be obtained from the following combinations\\
\begin{align} \label{eq:Gauge tt p-s Non-Int Finite T}
\begin{split}
    \text{(left poles \& right zeros):} \qquad\qquad\qquad\qquad\qquad \\
        i \omega = 2 \pi T_R \left\{ -\frac{\nu}{2} + n^z_R \right\} &+ 2 \pi T_L \left\{ \frac{\nu}{2} + n^p_L \right\}\,,\\
        i k = 2 \pi T_R \left\{ -\frac{\nu}{2} + n^z_R \right\} &- 2 \pi T_L \left\{ \frac{\nu}{2} + n^p_L \right\}\,,\\
    \text{(right poles \& left zeros):} \qquad\qquad\qquad\qquad\qquad \\
        i \omega = 2 \pi T_R \left\{ 1+\frac{\nu}{2} + n^p_R \right\} &+ 2 \pi T_L \left\{ 1 - \frac{\nu}{2} + n^z_L \right\}\,,\\
        i \omega = 2 \pi T_R \left\{ 1+\frac{\nu}{2} + n^p_R \right\} &- 2 \pi T_L \left\{ 1 - \frac{\nu}{2} + n^z_L \right\}\,.
\end{split}
\end{align}
Note that the remaining combination, (left poles \& left zeros) and (right \& right zeros) cannot produces the pole-skipping point.\\

It may be instructive to show the leading pole-skipping point to discuss the effect of rotation. One cn check that the first combination, \eqref{eq:Gauge tt p-s Non-Int Finite T}, produces the leading pole-skipping point $(\omega_{\text{leading}}, k_{\text{leading}})$k, in particular using \eqref{eq:TLTR OmegaT}, we find
\begin{equation} \label{LPSPHERM11 Gauge tt}
    (\omega_{\text{leading}}, k_{\text{leading}}) = \frac{2\pi i \,T_h}{1-\Omega^2}\bigg( \nu \, \Omega \,,\,\, \nu \bigg)\,,
\end{equation}
where the correction of rotation $\Omega$ appears only in frequency.

\paragraph{Pole-skipping in the extreme limit.}
The pole-skipping in the extreme limit $(T_L \, \ra \, 0)$ for the non integer case cannot be achieved simply by taking $T_L\,\ra\,0$ on \eqref{eq:Gauge tt p-s Non-Int Finite T} because \eqref{FPF1 Gauge tt}-\eqref{FPF2 Gauge tt} are ill-defined in the extreme limit.

Instead, we need to consider $G^R_{tt,0}$ of \eqref{eq:Gauge GR Non-Int Zero T} in order to discuss the pole-skipping in the extreme limit. From the structure of $G^R_{tt,0}$, one can find the two types of the `would-be' pole-skipping points, $(\omega_{(I)},\, k_{(I)})$ and $(\omega_{(II)},\, k_{(II)})$, as
\begin{align}\label{WBP1 Gauge tt}
\begin{split}
\nu > 1&:\,\,  \text{right poles} \,\,\,\&\,\,\, (\omega-k)^{\nu - 1} = 0 \quad \rightarrow\quad \omega_{(I)} = k_{(I)} = -i\,\pi T_R (2 + 2 n^{p}_{R} + \nu)\,,\\
\nu < 1&:\,\,  \text{right zeros} \,\,\,\&\,\,\, (\omega-k)^{\nu - 1} = \infty \hspace{0.22cm} \rightarrow\quad \omega_{(II)} = k_{(II)} = -i\,\pi T_R (2 n^{z}_{R} - \nu)\,,
\end{split} 
\end{align}
where right poles are from \eqref{FPF1 Gauge tt} and right zeros from \eqref{FPF2 Gauge tt}.

However, as same as the subsection \ref{subsec:Non half-integer nu case Fermi}, \eqref{WBP1 Gauge tt} may not be considered as the pole-skipping points. The vector Green's function $G^R_{tt,0}$ near the `would-be' pole-skipping points \eqref{WBP1 Gauge tt} can be expressed as
\begin{align}\label{CHECKINGPSW Gauge tt}
\begin{split}
G^R_{tt,0}(\omega_{(I)} + \delta \omega,\, k_{(I)} + \delta k)  &\,\,\approx\,\, F\left(\frac{\delta \omega}{\delta k}, \nu\right) \delta \omega^{\nu-2} \quad\rightarrow\quad
\begin{cases}
       \,\,\, 0 \qquad\,\, \left( \nu>2 \right) \,,\\
       \,\, \infty \qquad \left(1<\nu<2\right) \,, 
\end{cases}
\\
G^R_{tt,0}(\omega_{(II)} + \delta \omega,\, k_{(II)} + \delta k)  &\,\,\approx\,\, F\left(\frac{\delta \omega}{\delta k}, \nu\right) \delta \omega^{\nu-1} \quad\rightarrow\quad
\begin{cases}
       \,\,\, \infty \qquad \left(0<\nu<1\right) \,.
\end{cases}
\end{split} 
\end{align}
Therefore, \eqref{WBP1 Gauge tt} cannot be taken as the pole-skipping points since the Green's function $G^R_{tt,0}$ can be solely determined as $\infty$ or $0$.


\subsubsection{Integer $\nu$ case and pole-skipping in extreme limit} \label{sec:4.3.2}
\paragraph{Pole-skipping in the non-extremal case.}
For the integer $\nu$ case, because of the property of gamma function \eqref{eq:Gamma function ratio}, one can notice that the gamma functions in $G^R_{tt}$ of \eqref{eq:Gauge GR Int Finite T} can be expressed as
\begin{align} \label{GMFR12 Vector}
\begin{split}
&\frac{\Gamma(a_{L}+\nu-1)}{\Gamma(a_{L})}=\left(a_{L}+\nu-2\right)\times \left(a_{L}+\nu-3\right)\times \cdots \times a_{L} \,, \qquad \nu-1 \in \mathbb{Z}^{+} \,, \\
&\frac{\Gamma(b_{R}+\nu+1)}{\Gamma(b_{R})}= \left(b_{R}+\nu\right)\times \left(b_{R}+\nu-1\right) \times \cdots \times b_{R} \,, \qquad \qquad \nu+1 \in \mathbb{Z}^{+} \,,
\end{split}
\end{align}
These gamma functions and overall factor $(\omega-k)$ in the Green's function $G^R_{tt}$ give the following zeros
\begin{align} \label{zerosnhi1 Gauge tt}
\begin{split}
    \text{(left zeros):} \qquad & a_L = 1-\frac{\nu}{2} + \frac{\omega-k}{i\, 4 \pi T_L} = - n^z_L\,,\\
    \text{(right zeros):} \qquad & b_R = - \frac{\nu}{2} + \frac{\omega+k}{i\, 4 \pi T_R} = - n^z_R\,,
\end{split}
\end{align}
where integer $n^z_L=0,1,\cdots,\nu-2$ and $n^z_R=0,1,\cdots,\nu$. It differs from \eqref{FPF2 Gauge tt} in that $n^z_L$ and $n^z_R$ have upper bounds. Also, note that the gamma functions in $G^R_{tt}$ only produce zeros.

The poles of \eqref{eq:Gauge GR Int Finite T} are associated with the digamma functions therein, respectively, as
\begin{align} \label{polesnhi1 Gauge tt}
\begin{split}
    \text{(left poles):} \qquad & a_L+\nu-1 = \frac{\nu}{2} + \frac{\omega-k}{i\, 4 \pi T_L} = - n^p_L\,,\\
    \text{(right poles):} \qquad & b_R+\nu = \frac{\nu}{2} + \frac{\omega+k}{i\, 4 \pi T_R} = - n^p_R\,,
\end{split}
\end{align}
where integer $n^p_L=0,1,2,\cdots$.\\

As same as section \ref{subsec:Half-integer p-s Fermi}, the leading condition ($n_R^p=0$) may not only be irrelevant for our pole-skipping analysis but also affect the range for the zeros: $n_R^z$.
In this case, one can find the same condition as a ``zero" condition in the right zeros
(see $b_R=-n_R^z=-\nu$ from \eqref{zerosnhi1 Gauge tt}).
This implies that at ($n_R^z=\nu$) from right zeros or ($n_R^p=0$) from right poles, the Green's function \eqref{eq:Gauge GR Int Finite T} can be expressed as
\begin{equation} \label{DSRBE33 Gauge tt}
\begin{split}
{G}^R_{tt} (\omega, k)  \,\approx\, \left(b_{R}+\nu\right) \Psi\left({b}_R+\nu\right) \,\approx\, -1 \,,
\end{split}
\end{equation}
when $b_R + \nu=0$, which is a finite constant. The condition $b_R + \nu=0$ cannot be considered as zeros (or poles). Thus, the range for the right zeros is redefined as $n^z_R=0,1,\cdots, \nu-1$.\\

In summary, for the integer case, we have the (left/right) zeros \eqref{zerosnhi1 Gauge tt} and the (left/right) poles \eqref{polesnhi1 Gauge tt} with 
\begin{align}\label{HICRANGE Vector tt}
\begin{split}
n^z_L = 0,1,\cdots, \nu-2 \,,  \qquad \, & n^z_R = 0,1,\cdots, \nu-1 \,,\\
n^p_L = 0,1,2,\cdots \,, \hspace{0.8cm} \qquad
& n^p_R = 1,2,3,\cdots \,.
\end{split}
\end{align}
Based on the condition of $n$, we find the pole-skipping points as\\
\begin{align}\label{eq:Gauge tt p-s Int Finite T}
\begin{split}
\text{(left poles \& right zeros):} \qquad\qquad\qquad\qquad\qquad \\
i \omega = 2 \pi T_R \left\{ -\frac{\nu}{2} + n^z_R \right\} & + 2 \pi T_L \left\{ \frac{\nu}{2} + n^p_L \right\}\,,\\
i k = 2 \pi T_R \left\{ -\frac{\nu}{2} + n^z_R \right\} & - 2 \pi T_L \left\{ \frac{\nu}{2} + n^p_L \right\}\,,\\
\text{(right poles \& left zeros):} \qquad\qquad\qquad\qquad\qquad \\
i \omega = 2 \pi T_R \left\{ \frac{\nu}{2} +n^p_R \right\} & + 2 \pi T_L \left\{ 1-\frac{\nu}{2} + n^z_L \right\}\,,\\
i k = 2 \pi T_R \left\{ \frac{\nu}{2} + n^p_R \right\} & - 2 \pi T_L \left\{ 1-\frac{\nu}{2} + n^z_L \right\}\,,
\end{split}
\end{align}
which have the same structure with the non integer case \eqref{eq:Gauge tt p-s Non-Int Finite T}, if we just rewrite $n^p_R \ra n^p_R + 1$ with a range $n^p_R=0,1,2,\cdots$. However, notice that the functional form of the leading pole-skipping point is the same with \eqref{LPSPHERM11 Gauge tt}.\\

\paragraph{Pole-skipping in the extreme limit.}
Next, let us discuss the extreme limit of the pole-skipping for the integer case.
In order for this, we need to consider $G^R_{tt}$ in \eqref{eq:Gauge GR Int Zero T}
\begin{equation}\label{G0FORDD Vector tt}
    G^R_{tt,0} (\omega, k)  \approx \,\, \frac{\Gamma({b}_R+\nu+1)}{\Gamma({b}_R)} \left( \omega-k \right)^{\nu-1} \left[ \log (\omega-k) +\Psi\left({b}_R+\nu\right) \right]\,.
\end{equation}
We can find that the factor $\frac{\Gamma({b}_R+\nu+1)}{\Gamma({b}_R)} \left( \omega-k \right)^{\nu-1}$ only give the zeros as
\begin{align}\label{ZEL Vector tt}
\begin{split}
    \text{(left zeros):} \qquad & \omega - k = 0\,,\\
    \text{(right zeros):} \qquad & {b}_R = -\frac{\nu}{2} +\frac{\omega+k}{ i\, 4 \pi T_R} = - n^z_R\,,
\end{split}
\end{align}
where \eqref{HICRANGE Vector tt}.

Also, we can find the poles from the rest, $\left[ \log (\omega-k) +\Psi\left({b}_R+\nu\right) \right]$, as
\begin{align}\label{PEL Vector tt}
\begin{split}
    \text{(right poles):} \qquad & {b}_R+\nu = \frac{\nu}{2} + \frac{\omega+k}{i\, 4 \pi T_R} = - n^p_R\,,
\end{split}
\end{align}
where \eqref{HICRANGE Vector tt}. Notice that one may find the poles from $\log (\omega-k)$, $\omega-k=0$, however this cannot be considered as the poles since the structure of \eqref{G0FORDD Vector tt} shows that $\omega-k=0$ corresponds to the zeros: 
\begin{equation}\label{}
\begin{split}
\left( \omega-k \right)^{\nu}\log (\omega-k) \approx 0  \,,
\end{split}
\end{equation}
when $\omega \rightarrow k$.\\

Finally, combining the left zeros, $\omega-k=0$, with the right poles \eqref{PEL Vector tt}, we find the intersections of poles and zeros as
\begin{align}\label{PSPHIEXT Vector tt}
\text{(right poles \& left zeros):} \qquad
i \omega_{(0)} \, = \, i k_{(0)} = 2 \pi T_R \left\{ \frac{\nu}{2} + n^p_R \right\}\,,
\end{align}
Furthermore, as did in the non integer case \eqref{CHECKINGPSW Gauge tt}, we can expand the Green's function $G^R_{tt,0}$ near \eqref{PSPHIEXT Vector tt} and find
\begin{align}
G^R_{tt,0}(\omega_{(0)} + \delta \omega, k_{(0)} + \delta k)  &\quad\approx\quad F
\left(\frac{\delta \omega}{\delta k}, \nu\right)
\delta \omega^{\nu-2}  \,,
\end{align}
which implies that \eqref{PSPHIEXT Vector tt} can be considered as the pole-skipping point when $\nu=2$. Therefore, we can find the pole-skipping point even in the extreme limit for the integer case when $\nu=2$ as:
\begin{align}\label{answer111 Vector}
\left(\omega^{\text{ext}}\,,\, k^{\text{ext}}\right):= \left( \omega_{(0)}, k_{(0)} \right) \Big|_{\nu=2} = -2 \pi i T_R \left(1+n^p_R\right) \,,
\end{align}
in which its leading with $n^p_R=1$ is 
\begin{align}\label{eq:Zero T p-s Gauge}
\left( \omega^{\text{ext}}_\text{leading}\,,\, k^{\text{ext}}_\text{leading} \right) = -4 \pi i T_R \,.
\end{align}

\subsection{Pole-skipping points of $G^R_{rr}$}
\subsubsection{Non integer $\nu$ case}
\paragraph{Pole-skipping in the non-extremal case.}
For the Green's function $G^R_{rr}$ in \eqref{eq:Gauge GR Non-Int Finite T}, the poles and zeros are added from \eqref{FPF1 Gauge tt} and \eqref{FPF2 Gauge tt} by the overall factor $\frac{\omega-k}{\omega+k}$. The additional poles and zeros are
\begin{align} \label{FPF1 Gauge rr}
\begin{split}
    \text{(right pole 0):} \qquad & \omega+k \,=\,0\,,\\
    \text{(left zero 0):} \qquad & \omega-k \,=\, 0\,.
\end{split}
\end{align}
The pole-skipping points are at the intersects of the pole line and zero line in \eqref{FPF1 Gauge tt}, \eqref{FPF2 Gauge tt}, and \eqref{FPF1 Gauge rr}. Then, the additional pole-skipping points for $G^R_{rr}$ are obtained from those for $G^R_{tt}$ as the following combinations
\begin{align} \label{eq:Gauge rr p-s Non-Int Finite T}
\begin{split}
\text{(right poles \& left zeros 0):} \qquad &
i \omega = \,\, i k = 2 \pi T_R \left\{ 1 + \frac{\nu}{2} + n^p_R \right\}\,,
\\
\text{(right poles 0 \& left zeros):} \qquad &
i \omega = \, -i k = 2 \pi T_L \left\{ 1 - \frac{\nu}{2} + n^z_L \right\}\,,
\\
\text{(right poles 0 \& left zeros 0):} \qquad &
i \omega = \,\, i k = 0\,,
\end{split}
\end{align}
where the other pole-skipping points are the same as \eqref{eq:Gauge tt p-s Non-Int Finite T}.
\\

It is instructive to show the leading pole-skipping point to discuss the effect of rotation.
One can check that the first combination for $G^R_{rr}$ in \eqref{eq:Gauge tt p-s Non-Int Finite T} and \eqref{eq:Gauge rr p-s Non-Int Finite T}, produces the leading pole-skipping point $(\omega_\text{leading}, k_\text{leading})$ as same as one for $G^R_{tt}$ \eqref{LPSPHERM11 Gauge tt}.

\paragraph{Pole-skipping in the extreme limit.}
The pole-skipping in the extreme limit ($T_L \rightarrow 0$) for the non integer case cannot be achieved simply by taking $T_L \rightarrow 0$ on \eqref{eq:Gauge rr p-s Non-Int Finite T} because $a_L$ in \eqref{tildeab Gauge} are ill-defined in the extreme limit.

Instead, we need to consider $G^R_{rr,0}$ of \eqref{eq:Gauge GR Non-Int Zero T} in order to discuss the pole-skipping in the extreme limit. 
From the structure of $G^R_{rr,0}$, one can find the two types of the `would-be' pole-skipping points, ($\omega_{(I)},\, k_{(I)}$) and ($\omega_{(II)},\, k_{(II)}$), as
\begin{align}\label{WBP1 Gauge rr}
\begin{split}
\text{right poles} \,\,\,\&\,\,\, (\omega-k)^{\nu} = 0 \quad&\rightarrow\quad \omega_{(I)} = k_{(I)} = - i \pi T_R (2 + 2 n^{p}_{R} + \nu)\,,\\
\text{right pole 0} \,\,\,\&\,\,\, (\omega-k)^{\nu} = 0 \quad&\rightarrow\quad \omega_{(II)} = k_{(II)} = 0\,,
\end{split} 
\end{align}
where right poles are from \eqref{FPF1 Gauge tt} and \eqref{FPF1 Gauge rr}. Note that the overall factor $(\omega-k)^{\nu}$ only can make zeros because of the positive $\nu$. This is a difference with those for $G^R_{tt,0}$ in \eqref{WBP1 Gauge tt}.

However, as same as the subsection \ref{subsec:Non half-integer nu case Fermi}, \eqref{WBP1 Gauge rr} may not be considered as the pole-skipping points. The vector Green's function $G^R_{rr,0}$ near the `would-be' pole-skipping points \eqref{WBP1 Gauge rr} can be expressed as
\begin{align}\label{CHECKINGPSW Gauge rr}
\begin{split}
G^R_{rr,0}(\omega_{(I)} + \delta \omega, k_{(I)} + \delta k)  &\,\,\approx\,\, F\left(\frac{\delta \omega}{\delta k}, \nu\right) \delta \omega^{\nu-1} \quad\rightarrow\quad
\begin{cases}
       \,\,\, 0 \qquad\,\, \left(\nu>1\right) \,,\\
       \,\, \infty \qquad \left(0<\nu<1\right) \,,
\end{cases}
\\
G^R_{rr,0}(\omega_{(II)} + \delta \omega, k_{(II)} + \delta k)  &\,\,\approx\,\, F\left(\frac{\delta \omega}{\delta k}, \nu\right) \delta \omega^{\nu-1} \quad\rightarrow\quad
\begin{cases}
       \,\,\, 0 \qquad\,\, \left(\nu>1\right) \,,\\
       \,\, \infty \qquad \left(0<\nu<1\right) \,.
\end{cases}
\end{split} 
\end{align}
Therefore, \eqref{WBP1 Gauge rr} cannot be taken as the pole-skipping points since the Green's function $G^R_{rr,0}$ can be solely determined as $\infty$ or $0$.

\subsubsection{Integer $\nu$ case and pole-skipping in extreme limit}
\paragraph{Pole-skipping in the non-extremal case.}
For the integer $\nu$ case, the Green's function $G^{R}_{rr}$ of \eqref{eq:Gauge GR Int Finite T} also have the overall factor $\frac{\omega-k}{\omega+k}$. The additional pole and zeros are found by the factor as
\begin{subequations} \label{zerosnhi1 Gauge}
\begin{align}
    \text{(left zero 0):} \qquad & \omega-k=0  \,,\\
    \text{(right pole 0):} \qquad & \omega+k=0 \,,
\end{align}
\end{subequations}
than \eqref{zerosnhi1 Gauge tt} and \eqref{polesnhi1 Gauge tt} of $G^R_{tt,0}$. The condition of $n$ is also same with \eqref{HICRANGE Vector tt} as
\begin{align*}
\begin{split}
n^z_L = 0,1,\cdots, \nu-2 \,,  \qquad \, & n^z_R = 0,1,\cdots, \nu-1 \,,\\
n^p_L = 0,1,2,\cdots \,, \hspace{0.8cm} \qquad
& n^p_R = 1,2,3,\cdots \,.
\end{split}
\end{align*}
Therefore, the additional intersection points are
\begin{align} \label{eq:Gauge rr p-s Int Finite T}
\begin{split}
\text{(right poles \& left zeros 0):} \qquad\qquad & i \omega = \,\, i k = 2 \pi T_R \left\{ \frac{\nu}{2} + n^p_R \right\}\,,
\\
\text{(right poles 0 \& left zeros):} \qquad\qquad & i \omega = \, -i k = 2 \pi T_L \left\{ 1 - \frac{\nu}{2} + n^z_L \right\}\,,
\\
\text{(right poles 0 \& left zeros 0):} \qquad\qquad & i \omega = \,\, i k = 0\,,
\end{split}
\end{align}
where the other points are \eqref{eq:Gauge tt p-s Int Finite T} with \eqref{HICRANGE Vector tt}. However, it differs from those of $G^R_{tt}$ that many points of $G^R_{rr}$ in \eqref{eq:Gauge tt p-s Int Finite T} and \eqref{eq:Gauge rr p-s Int Finite T} are not pole-skipping points. For instance, one can see the expansion of Green's function near the intersection point $(\omega_\star, k_\star)$ for (right pole \& left zeros) in \eqref{eq:Gauge rr p-s Int Finite T} at $\nu=2$, $n^z_L=0$, $n^p_R=1$ such as
\begin{equation}
    G^R_{rr}(\omega_\star + \delta \omega, k_\star + \delta k) \, \approx \, F\left(\frac{\delta \omega}{\delta k}\right) \delta \omega + \mathcal{O}(\delta \omega^2)\,.
\end{equation}
This kind of exceptions which are intersections but no pole-skipping points is caused by meeting different three (pole or zero) lines at that point.
By the way, even though these exceptions, the functional form of the leading pole-skipping point is also the same with \eqref{LPSPHERM11 Gauge tt}.\\

\paragraph{Pole-skipping in the extreme limit.}
For discussing the extreme limit of the pole-skipping with the integer $\nu$, we need to consider $G^R_{rr,0}$ in \eqref{eq:Gauge GR Int Zero T}
\begin{equation}\label{G0FORDD Vector rr}
\begin{split}
    G^R_{rr,0} (\omega, k)  \approx \,\, & \frac{\Gamma({b}_R+\nu+1)}{\Gamma({b}_R)} \frac{ \left( \omega-k \right)^{\nu} }{ \omega + k } \left[ \log (\omega-k) +\Psi\left({b}_R+\nu\right) \right]\,.
\end{split}
\end{equation}
In this case, the overall factor $\frac{(\omega-k)^\nu}{\omega+k}$ makes difference from $G^R_{tt,0}$. Therefore, there is an additional right poles as
\begin{align}\label{PEL Vector rr}
    \text{(right poles 0):} \qquad & \omega + k = 0\,.
\end{align}
than \eqref{ZEL Vector tt} and \eqref{PEL Vector tt}.\\

Considering zeros and poles in \eqref{ZEL Vector tt}, \eqref{PEL Vector tt}, and \eqref{PEL Vector rr} for $G^R_{rr,0}$, we can find the all the possible intersections as
\begin{align}\label{PSPHIEXT Vector rr}
\begin{split}
\text{(right poles \& left zeros 0):} \qquad
& i \omega_{(I)} = \,\, i k_{(I)} = 2 \pi T_R \left\{ \frac{\nu}{2} + n^p_R \right\}\,,
\\
\text{(right poles 0 \& left zeros 0):} \qquad
& i \omega_{(II)} = \,\, i k_{(II)} = 0\,,
\end{split}
\end{align}
where \eqref{HICRANGE Vector tt}.
Also, we can expand the Green's function $G^R_{rr,0}$ near \eqref{PSPHIEXT Vector rr} and find
\begin{align}
G^R_0(\omega_{(I)} + \delta \omega, k_{(I)} + \delta k)  &\quad\approx\quad F
\left(\frac{\delta \omega}{\delta k}, \nu\right)
\delta \omega^{\nu-1}  \,,\\
G^R_0(\omega_{(II)} + \delta \omega, k_{(II)} + \delta k)  &\quad\approx\quad F
\left(\frac{\delta \omega}{\delta k}, \nu\right)
\delta \omega^{\nu-1}  \,,
\end{align}
which implies that \eqref{PSPHIEXT Vector rr} could be considered as the pole-skipping point when $\nu=1$.
However, we can not find any pole-skipping point because the Green's function $G^R_{rr,0}$ is computed on the assumption as integer $\nu\geq2$ in \eqref{eq:Int nu range vector}.

%
\section{Conclusion}\label{sec:Conclusions}

We have studied the pole-skipping points in the momentum space ($\omega, k$) of Green's functions in rotating BTZ background.
In particular, inspired by the analysis of scalar field case~\cite{Natsuume:2020snz}, we have investigated the case of fermionic and vector fields, which allow the analytic calculations of holographic Green's function.
We have shown the full tower of pole-skipping points including the effect of rotation: we summarized our results in the Table. \ref{TB111}.
\begin{table}[]
\begin{tabular}{| C{2.68cm} | C{5.5cm} | C{5.5cm}  |}
\hline
&   \textbf{Non-extremal black holes}    &      \textbf{Extremal black holes}  \\ 
 \hline
 \hline
    Fermionic fields ($s=1/2$)      &  Non half-integer case \eqref{eq:Fermionic p-s Non-Int Finite T},   Half-integer case \eqref{eq:Fermionic p-s Int Finite T222}.          &  Half-integer case with $\nu = 3/2$ \eqref{answer111}. \\
\hline
     Vector fields ($s=1$)    &   Non integer case \eqref{eq:Gauge tt p-s Non-Int Finite T} \& \eqref{eq:Gauge rr p-s Non-Int Finite T}, Integer case \eqref{eq:Gauge tt p-s Int Finite T} \& \eqref{eq:Gauge rr p-s Int Finite T}.    &        
Integer case with $\nu=2$ \eqref{answer111 Vector}. \\
 \hline
\end{tabular}
\caption{Summary of pole-skipping points of fermionic and vector fields in rotating BTZ.}\label{TB111}
\end{table}

Our study resulted in two main findings.
To begin with, for the \textit{non-extremal} black holes, we have shown that the leading pole-skipping point in the previous literature ($\Omega=0$) can be generalized as 
\begin{equation} \label{eq:Conclusion 1}
    (\omega_{\text{leading}}, k_{\text{leading}})=\frac{2\pi i T_h}{1-\Omega^2}\bigg( s-1 + \nu \Omega\,,~ (s-1)\Omega + \nu \bigg)\,,
\end{equation}
where we find the correction of the rotation $\Omega$. Here, $T_h$ is the Hawking temperature, $s$ the spin of fields, and $\nu$ is defined as the difference of conformal dimensions $\nu := \frac{1}{2}(\Delta_+ - \Delta_-)$.
We also confirmed our leading pole-skipping result \eqref{eq:Conclusion 1} using another method, the near horizon analysis (see appendix \ref{sec:Near-horizon analysis}): note that we also examined the cases of $s=3/2, 2$ as well as $s=1/2, 1$.\footnote{We have also discussed the pole-skipping in the comoving frame in appendix \ref{sec:Near-horizon analysis}.}

Secondly, we identified that in the case of \textit{extremal} black holes, a pole-skipping point can only exist when the condition
\begin{equation}\label{MAR2}
    \nu = s+1 \,,
\end{equation}
is satisfied. The corresponding leading pole-skipping point is given by:
\begin{equation}\label{MAR3}
    \omega^{\text{ext}}_{\text{leading}} = k^{\text{ext}}_{\text{leading}} = 
    - 2\pi i T_R (s+1)\,.
\end{equation}
This result cannot be obtained by simply taking the extreme limit ($T_h\rightarrow 0$ and $\Omega\rightarrow 1$) of the non-extremal black holes \eqref{eq:Conclusion 1}.\\

Since the near-horizon analysis method is not applicable to extremal black holes, as explained in appendix \ref{sec:Near-horizon analysis}, it is imperative to develop alternative methods to validate \eqref{MAR2}-\eqref{MAR3}. Recently, a new method has been proposed in \cite{Natsuume:2023lzy}, which shows promise in addressing this issue. We leave this subject as future work and hope to address it in the near future.

\acknowledgments

We would like to thank {Yongjun Ahn, Kyung-Sun Lee, Makoto Natsuume, Mitsuhiro Nishida} for valuable discussions and correspondence.
This work was supported by the Basic Science Research Program through the National Research Foundation of Korea (NRF) funded by the Ministry of Science, ICT $\&$ Future Planning (NRF-2021R1A2C1006791) and GIST Research Institute(GRI) grant funded by the GIST in 2023. This work was also supported by Creation
of the Quantum Information Science R$\&$D Ecosystem (Grant No. 2022M3H3A106307411)
through the National Research Foundation of Korea (NRF) funded by the Korean government (Ministry of Science and ICT).
H.-S Jeong acknowledges the support of the Spanish MINECO ``Centro de Excelencia Severo Ochoa'' Programme under grant SEV-2012-0249. This work is supported through the grants CEX2020-001007-S and PID2021-123017NB-I00, funded by MCIN/AEI/10.13039/501100011033 and by ERDF A way of making Europe.
C.-W Ji and H.-S Jeong contributed equally to this paper and should be considered co-first authors.

\appendix
%
\section{Near-horizon analysis} \label{sec:Near-horizon analysis}
We suggested a general form of the leading pole-skipping points in \eqref{eq:Conclusion 1} with respect to spin $s$ at finite temperature. The general form is obtained by the analytic Green's function for spin $s=0,\, \frac{1}{2},\, 1$. In this appendix, the leading pole-skipping is double-checked by the near-horizon analysis for $s=\frac{1}{2},\, 1$. The scalar field case($s=0$) was already checked in \cite{Natsuume:2020snz}. Furthermore, although the analytic Green's function is not computed, the leading pole-skipping form \eqref{eq:Conclusion 1} can be checked by near-horizon analysis. Thus, we also compute the leading pole-skipping points for the spin $s=\frac{3}{2},\, 2$.

However, the near-horizon analysis is not useful at the extreme limit. For the scalar field, the horizon $r=r_+$ is a regular singularity at finite temperature but becomes an irregular singularity at the extreme limit. Then, the Frobenius method cannot be used \cite{Natsuume:2020snz}. Similarly, the near-horizon analysis is not also valid at the extreme limit for higher spins. Thus, we analyze the pole-skippings by using analytic solutions in sections \ref{sec:Fermionic fields} and \ref{sec:Vector fields}.

\subsection{Fermionic/Vector fields ($s=1/2, 1$)}\label{subsec:Fermionic/Vector fields}
\paragraph{Fermionic field}
The Dirac equation of $\chi_{1,2}$ in \eqref{eq:Dirac Eqn chi12 a} is given by
\begin{subequations} \notag
\begin{align}
    2(1-z)\sqrt{z}\partial_z \chi_1 - i\left( \frac{k_T}{\sqrt{z}} + k_X \sqrt{z} \right) \chi_1 = \left( m_f-\frac{1}{2}+i(k_T+k_X) \right)\chi_2\,.
\end{align}
\end{subequations}
From the decoupled second order differential equation, the leading behaviors of the incoming solution $\chi_{1,2}$ are found as
\begin{equation}
    \chi_1 \sim z^{\frac{1}{2}-i \lambda}\,, \qquad\qquad \chi_2 \sim z^{-i \lambda}\,, 
\end{equation}
where $\lambda := \frac{\omega - \Omega k}{4 \pi T_h}$ at the horizon, by the relations in \eqref{eq:TLTR OmegaT} and \eqref{eq:kTkX}. According to these leading behaviors, the series expansion as
\begin{equation}
    \chi_1 = z^{\frac{1}{2}-i \lambda} \sum^{\infty}_{j=0}\chi_1^{(j)} z^j\,, \qquad \qquad \chi_2 = z^{-i \lambda} \sum^{\infty}_{j=0}\chi_2^{(j)} z^j\,, 
\end{equation}
are substituted into the equation \eqref{eq:Dirac Eqn chi12 a} and the horizon expansion of this equation becomes
\begin{equation} \label{eq:Fermion horizon series equation}
    z^{-i \lambda} \sum^{
    \infty}_{j=0} S^{(j)}(\omega, k)z^j =0\,.
\end{equation}
The series coefficients $S^{(j)}$ are computed such as 
\begin{equation}
    S^{(0)}=
    \left( 1-i\frac{\omega - k \Omega}{\pi T_h} \right)\chi_1^{(0)}
    +\left(\frac{1}{2}-m_f -i (1-\Omega) \frac{\omega+k}{2\pi T_h} \right)\chi_2^{(0)}
    =0\,.
\end{equation}
By that the coefficients of $\chi_{1,2}^{(0)}$ vanish, the leading pole-skipping points are computed as
\begin{equation}
    (\omega, k)=\frac{2\pi i T_h}{1-\Omega^2}\bigg( -\frac{1}{2} + m_f \Omega\,,~ -\frac{1}{2}\Omega + m_f \bigg)\,.
\end{equation}
This is consistent with the one of non-integer case at finite temperatures in \eqref{eq:Fermionic p-s Non-Int Finite T}.

\paragraph{Gauge field}
For simplicity, comoving coordinates and incoming Eddington-Finkelstein coordinates are used. In the comoving coordinates $(t,r,\phi)$, the metric is
\begin{equation}
    {\td s}^2 =-f(r) {\td t}^2+\frac{{\td r}^2}{f(r)}+r^2\left(\frac{r_-}{r_+}\frac{r^2-r_+^2}{r^2}\td t + \td \phi\right)^2\,,
\end{equation}
where
\begin{equation}
    \phi=\varphi-\frac{r_-}{r_+}t\,.
\end{equation}
The incoming Eddington-Finkelstein coordinates $(v,r,\phi)$ are obtained with $v=t+r_*$ where
\begin{equation}
    r_*=\frac{r_+}{2(r_+^2-r_-^2)}\log\frac{\sqrt{r^2-r_-^2}-\sqrt{r_+^2-r_-^2}}{\sqrt{r^2-r_-^2}+\sqrt{r_+^2-r_-^2}}\,.
\end{equation}
The metric becomes
\begin{equation} \label{eq:metric incoming E-F}
\begin{split}
    ds^2 &= -\frac{(r^2-r_+^2)(r_+^2-r_-^2)}{r_+^2} dv^2 + 2 \frac{r}{r_+}\sqrt{\frac{r_+^2 - r_-^2}{r^2 - r_-^2}} dv dr +  \frac{2 r_-(r^2 - r_+^2)}{r_+}dv d\phi \\ &\hspace{6cm} - \frac{2 r r_{-}}{\sqrt{(r^2 - r_-^2)(r_+^2 - r_-^2)}}dr d\phi + r^2 d\phi^2\,.
\end{split}
\end{equation}
The momentum variables $(\omega,k)$ in the coordinates $(t,r,\varphi)$ have a relation with the momentum variables $(\tilde{\omega},\tilde{k})$ in the incoming Eddington-Finkelstein coordinates $(v,r,\phi)$ as
\begin{equation} \label{eq:wk To E-F}
    \tilde{\omega}=\omega-\frac{r_-}{r_+}k\,, \qquad \qquad \tilde{k} = k\,.
\end{equation}
By substituting series expansions for $A_\lambda$ such as
\begin{equation}
    A_\lambda = \sum^{\infty}_{j=0}A_\lambda^{(j)} r^j\,,
\end{equation}
the massive Maxwell equations in \eqref{eq:Gauge eqn} becomes
\begin{equation}
    \sum^{
    \infty}_{j=0} S_\lambda^{(j)}(\tilde{\omega}, \tilde{k})r^j =0\,.
\end{equation}
The series coefficients $S_\lambda^{(j)}$ are computed such as
\begin{equation} \label{eq:Gauge NHA series eqn}
    S_v^{(0)} =
    \left[\frac{ 1 - \Omega^2 }{4 \pi^2 T_h^2} \bigg( \tilde{k} \tilde{\omega} \Omega - \tilde{k}^2 \left( 1-\Omega ^2 \right) \bigg) - m_A^2\right] \mathfrak{A}_v^{(0)}
    + \tilde{\omega} \bigg[ \tilde{\omega} \mathfrak{A}_r^{(0)}
    + \frac{m_A^2}{\tilde{k}} A_x^{(0)}
    - i \mathfrak{A}_v^{(1)} \bigg]
    = 0 \,,
\end{equation}
where
\begin{equation}
    \mathfrak{A}_v^{(j)}=\left(A_v^{(j)} + \frac{\tilde{\omega}}{\tilde{k}} A_x^{(j)} \right)\,, \qquad \qquad
    \mathfrak{A}_r^{(j)}=\left(A_r^{(j)} - \frac{j+1}{i\tilde{k}} A_x^{(j+1)} \right)\,.
\end{equation}
By that the all the coefficients of $A_{\lambda}$ vanish in \eqref{eq:Gauge NHA series eqn}, the leading pole-skipping points are found as
\begin{equation}
    (\tilde{\omega}, \tilde{k}) = \left(0\,, ~ \pm \frac{2 \pi i T_h m_A}{1 - \Omega^2} \right)\,.
\end{equation}
This point living in incoming Eddington-Finkelstein coordinates can be transformed to the Schwarzschild coordinates by the relation \eqref{eq:wk To E-F}. The leading pole-skipping points become
\begin{equation}
    (\omega, k) = \frac{2\pi i T_h}{1-\Omega^2}\left(\pm m_A \Omega, \pm m_A \right)\,.
\end{equation}
This result is consistent with the conclusion \eqref{eq:Conclusion 1} for $s=1$, $\nu=m_A$.

\subsection{Rarita-Schwinger fields ($s=3/2$)}\label{subsec:Rarita-Schwinger fields}
There is a near horizon analysis of Rarita-Schwinger field on static BTZ in \cite{Ceplak:2021efc}. This is a development on rotating BTZ. The massive Rarita-Schwinger field $\psi_N$ satisfies the equation given by
\begin{equation}\label{eq:Rarita eqn}
    \Gamma^{MNP} \nabla_N \psi_P - m_R \Gamma^{MN} \psi_N =0\,,
\end{equation}
where the covariant derivative is given by
\begin{equation}
    \nabla_M \psi_P = \partial_M \psi_P - \tilde{\Gamma}^N_{MP}\psi_N + \frac{1}{4} (\omega_{ab})_M \Gamma^{ab} \psi_P\,.
\end{equation}
As same as in section \ref{sec:Fermionic fields}, the capital letters such as $M, N, \cdots$ are for the curved coordinates and  the small letters such as $a, b, \cdots$ are for the tangent coordinates. The equation in \eqref{eq:Rarita eqn} becomes
\begin{align}\label{eq:Rarita eqn 2}
\left(\slashed \nabla + m_R\right) \psi_N = 0\,,
\end{align}
by two additional constraints
\begin{equation} \label{eq:Rarita const a}
\Gamma^M\, \psi_M =0\,, \qquad \qquad
\nabla^M \, \psi_M  = 0\,,
\end{equation}
where $\slashed \nabla := \Gamma^{M} \, \nabla_M$, $\Gamma^M=\Gamma^a e^M_a$, and $e^M_a$ vielbeins in \eqref{eq:VielBein}. In the coordinate $(T,\rho,X)$ in \eqref{eq:metric TrhoX}, the solution ansatz is given by
\begin{equation}
    \psi_M=\begin{pmatrix}
    \psi_M^{(+,+)} \\
    \psi_M^{(+,-)} \\
    \psi_M^{(-,+)} \\
    \psi_M^{(-,-)}
    \end{pmatrix}
    e^{-i k_T T + i k_X X} \,,
\end{equation}
where $M=T,\rho,X$. The vector components of Rarita-Schwinger field $\Psi_{M}$ are defined by eigenvectors of the matrices $\Gamma^{\underline{\rho}}$ and $\Gamma^{(2)}:=\Gamma^{\underline{TX}}$ as
\begin{equation}
    \Gamma^{\underline{\rho}}\psi_M^{(\pm,\alpha_2)}=\pm \psi_M^{(\pm,\alpha_2)}\,, \qquad 
    \Gamma^{(2)}\psi_M^{(\alpha_1,\pm)}=\pm \psi_M^{(\alpha_1,\pm)}\,,
\end{equation}
where the gamma matrices $\Gamma^a$ are chosen as
\begin{equation}
    \Gamma^{\underline{T}}=\begin{pmatrix}
    0 & 0 & 0 & 1 \\
    0 & 0 & -1 & 0 \\
    0 & 1 & 0 & 0 \\
    -1 & 0 & 0 & 0
    \end{pmatrix}\,,\quad
    \Gamma^{\underline{\rho}}=\begin{pmatrix}
    1 & 0 & 0 & 0 \\
    0 & 1 & 0 & 0 \\
    0 & 0 & -1 & 0 \\
    0 & 0 & 0 & -1
    \end{pmatrix}\,,\quad
    \Gamma^{\underline{X}}=\begin{pmatrix}
    0 & 0 & 0 & 1 \\
    0 & 0 & 1 & 0 \\
    0 & 1 & 0 & 0 \\
    1 & 0 & 0 & 0
    \end{pmatrix}\,.
\end{equation}
The $\psi_X$ can be decoupled from $\psi_T$ and $\psi_\rho$ by using the constraint \eqref{eq:Rarita const a}. The system of two field $\psi_T$ and $\psi_\rho$ is also decoupled by two matrices $\Gamma^{\underline{\rho}}$ and $\Gamma^{(2)}$. One decoupled system is for $\psi_T^{(\pm,\pm)}$ and $\psi_\rho^{(\pm,\pm)}$, and the other one is for $\psi_T^{(\pm,\mp)}$ and $\psi_\rho^{(\pm,\mp)}$. The two subsystems have two relations such as $k\ra -k$ and $ \Omega \ra - \Omega $. Therefore, it is enough to analyze the equations of $\psi_T^{(\pm,\pm)}$ and $\psi_\rho^{(\pm,\pm)}$:
\begin{align} \label{eq:Rarita eqn 3}
\begin{split}
    & \left[ 2(1-z)\sqrt{z} \partial_z + m_R - \frac{1-z}{2 \sqrt{z}} \right] \psi_T^{(+,+)}\\
    &  \hspace{1cm} + \frac{i\sqrt{1-z}}{2\pi T_h}\left[ k\left(1 +\frac{\Omega}{\sqrt{z}}\right) -\omega \left(\frac{1}{\sqrt{z}}+\Omega \right) \right] \Gamma^{\underline{T}} \psi_T^{(-,-)}
    - \frac{1}{\sqrt{1-z}}\Gamma^{\underline{T}} \psi_\rho^{(-,-)} =0\,,
    \\
    & \left[ 2(1-z)\sqrt{z} \partial_z - m_R - \frac{1-z}{2 \sqrt{z}} \right] \psi_T^{(-,-)}\\
    &  \hspace{1cm} + \frac{i\sqrt{1-z}}{2\pi T_h}\left[ k\left(1 - \frac{\Omega}{\sqrt{z}}\right) + \omega \left(\frac{1}{\sqrt{z}} - \Omega \right) \right] \Gamma^{\underline{T}} \psi_T^{(+,+)}
    + \frac{1}{\sqrt{1-z}}\Gamma^{\underline{T}} \psi_\rho^{(+,+)} =0\,,
    \\
    & \left[ 2(1-z)\sqrt{z} \partial_z + m_R + \frac{1 + 3z}{2 \sqrt{z}} \right] \psi_\rho^{(+,+)}\\
    &  \hspace{1cm} + \frac{i\sqrt{1-z}}{2\pi T_h}\left[ k\left(1 + \frac{\Omega}{\sqrt{z}}\right) - \omega \left(\frac{1}{\sqrt{z}} + \Omega \right) \right] \Gamma^{\underline{T}} \psi_\rho^{(-,-)}
    - \frac{(1-z)^{\frac{3}{2}}}{z}\Gamma^{\underline{T}} \psi_T^{(-,-)} =0\,,
    \\
    & \left[ 2(1-z)\sqrt{z} \partial_z - m_R + \frac{1 + 3z}{2 \sqrt{z}} \right] \psi_\rho^{(-,-)}\\
    &  \hspace{1cm} + \frac{i\sqrt{1-z}}{2\pi T_h}\left[ k\left(1 - \frac{\Omega}{\sqrt{z}}\right) + \omega \left(\frac{1}{\sqrt{z}} - \Omega \right) \right] \Gamma^{\underline{T}} \psi_\rho^{(+,+)}
    + \frac{(1-z)^{\frac{3}{2}}}{z}\Gamma^{\underline{T}} \psi_T^{(+,+)} =0\,.
\end{split}
\end{align}
By substituting series expansions for all components of $\psi_T^{(\pm,\pm)}$ and $\psi_\rho^{(\pm,\pm)}$ such as
\begin{align} \label{eq:Rarita series sol ansatz}
    \psi_T^{(\pm,\pm)} = z^{\alpha+\frac{1}{2}} \sum^{\infty}_{j=0} \psi _T^{(\pm,\pm)(j)} z^{j/2} \,, \qquad
    \psi_\rho^{(\pm,\pm)} = z^\alpha \sum^{\infty}_{j=0} \psi _\rho^{(\pm,\pm)(j)} z^{j/2} \,, 
\end{align}
the horizon expansions of the equations in \eqref{eq:Rarita eqn 3} become
\begin{equation}
\begin{split}
    z^{\alpha+\frac{1}{2}} \sum^{\infty}_{j=-2} S_1^{(j)}(\omega, k) z^{j/2} = 0 \,, \qquad  & z^{\alpha+\frac{1}{2}} \sum^{\infty}_{j=-2} S_2^{(j)}(\omega, k) z^{j/2} =0 \,, \\
    z^{\alpha} \sum^{\infty}_{j=-2} S_3^{(j)}(\omega, k) z^{j/2} =0 \,, \qquad  & z^{\alpha} \sum^{\infty}_{j=-2} S_4^{(j)}(\omega, k) z^{j/2} =0 \,.
\end{split}
\end{equation}
By setting the series coefficients $S_i^{(j)}$ with negative $j$ to zero, two different incoming solutions are found as
\begin{equation}
    \alpha_1 = - \frac{3}{4} - i \lambda\,, \qquad \alpha_2 = \frac{1}{4} - i \lambda \,,
\end{equation}
where $\lambda := \frac{\omega - \Omega k}{4 \pi T_h}$. For one incoming solution of $\alpha_1$, the set of equations $S_i^{(0)}=0$ where $i=1,2,3,4$ in \eqref{eq:Rarita eqn 3} rearranges as a matrix form as
\begin{equation} \label{eq:Rarita series equation}
    (2\pi i T_h - 2 \omega +2 k \Omega)
    \begin{pmatrix}
    \psi_T^{(+,+)(1)}\\  \psi_T^{(-,-)(1)}\\
    \psi_\rho^{(+,+)(1)}\\ \psi_\rho^{(-,-)(1)}
    \end{pmatrix}
    -
    (k + 2\pi i T_h m_R - \omega \Omega)
    \begin{pmatrix}
    \psi_\rho^{(+,+)(0)}\\ \psi_\rho^{(+,+)(0)}\\ \psi_\rho^{(+,+)(0)}\\  \psi_\rho^{(+,+)(0)}
    \end{pmatrix}=0\,,
\end{equation}
where the parameter $\psi_\rho^{(+,+)(0)}$ also determines all the other series coefficients of $\psi_T^{(\pm,\pm)}$ and $\psi_\rho^{(\pm,\pm)}$ in \eqref{eq:Rarita series sol ansatz}. By that the coefficients of $\psi_{T,\rho}^{(\pm,\pm)(1)}$ and $\psi_{\rho}^{(\pm,\pm)(0)}$ vanish in \eqref{eq:Rarita series equation}, the pole-skipping points are computed as
\begin{equation}
    (\omega, k)=\frac{2\pi i T_h}{1-\Omega^2}\bigg( \frac{1}{2} - m_R \Omega\,,~ \frac{1}{2}\Omega - m_R \bigg)\,.
\end{equation}
For another incoming solutions of $\alpha_2$, by the same way, the pole-skipping is computed as 
\begin{equation}
    (\omega, k)=\frac{2\pi i T_h}{1-\Omega^2}\bigg( -\frac{3}{2} - m_R \Omega\,,~ -\frac{3}{2}\Omega - m_R \bigg)\,.
\end{equation}
In the other decoupled subsystem for $\psi_T^{(\pm,\mp)}$ and $\psi_\rho^{(\pm,\mp)}$, the pole-skipping points are computed as
\begin{align}
\begin{split}
    (\omega, k)=\frac{2\pi i T_h}{1-\Omega^2}\bigg( \frac{1}{2} + m_R \Omega\,,~ \frac{1}{2}\Omega + m_R \bigg)\,, \qquad
    & \text{for } \alpha_1, \\
    (\omega, k)=\frac{2\pi i T_h}{1-\Omega^2}\bigg( -\frac{3}{2} + m_R \Omega\,,~ -\frac{3}{2}\Omega + m_R \bigg)\,, \qquad
    & \text{for } \alpha_2.
\end{split}
\end{align}
As a result, the leading pole-skipping points are
\begin{equation}\label{eq:Rarita Leading p-s}
    (\omega, k)=\frac{2\pi i T_h}{1-\Omega^2}\bigg( \frac{1}{2} \pm m_R \Omega\,,~ \frac{1}{2}\Omega \pm m_R \bigg)\,.
\end{equation}

Near the AdS boundary $z\ra 1$, the asymptotic behavior of the Rarita-schwinger field $\psi_M$ is found as
\begin{equation}
    \begin{split}
        \psi_T^{(+,+)} \sim A (1-z)^{\frac{1-m_R}{2}} + B (1-z)^{1+\frac{m_R}{2}} \qquad \psi_T^{(-,-)} \sim C (1-z)^{1-\frac{m_R}{2}} + D (1-z)^{\frac{1+m_R}{2}} \\
        \psi_\rho^{(+,+)} \sim E (1-z)^{\frac{3-m_R}{2}} + F (1-z)^{1+\frac{m_R}{2}} \qquad \psi_\rho^{(-,-)} \sim G (1-z)^{1-\frac{m_R}{2}} + H (1-z)^{\frac{3+m_R}{2}}
    \end{split}
\end{equation}
where $A$ and $D$ can determine all the other boundary series coefficients. The subsystem for $\psi_T^{(\pm,\mp)}$ and $\psi_\rho^{(\pm,\mp)}$ also has the same boundary behavior. If the Green's function is defined such as
\begin{equation}
    \tilde{G}^R = i \frac{D}{A}\,,
\end{equation}
as same as spin $1/2$ case in \eqref{eq:Gtilde}, the conformal dimension is also found as $\Delta_{\pm}=1\pm m_R$. Substituting $\nu=(\Delta_+ - \Delta_-)/2=m_R$ to \eqref{eq:Rarita Leading p-s}, the leading pole-skipping point is consistent with the conclusion in \eqref{eq:Conclusion 1}.

\subsection{Energy-density correlator ($s=2$)}\label{subsec:Energy-density correlator}
This is a brief review of leading pole-skipping points for metric perturbations around rotating BTZ in \eqref{eq:metric}, based on \cite{Liu:2020yaf}. For simplicity, the incoming Eddington-Finkelstein coordinates $(v,r,x)$ in \eqref{eq:metric incoming E-F} are used.
The Einstein equation for metric is given by
\begin{equation}
    R_{ab}-\frac{1}{2}R g_{ab} - g_{ab} = 0\,,
\end{equation}
where the metric $g_{ab}=\bar{g}_{ab}+h_{ab}$. The $\bar{g}_{ab}$ is a vacuum solution of Einstein equation and $h_{ab}$ is a linearized perturbation. By using $h_{ab}$, the Einstein equation can be written as
\begin{equation} \label{eq:Energy density eqn}
    E_{ab}:=\nabla^c\nabla_{(a}h_{b)c}-\frac{1}{2}\nabla^c \nabla_c h_{ab} -\frac{1}{2}\nabla_a \nabla_b h -\frac{1}{2} \bar{g}_{ab} \nabla^c \nabla^d h_{cd} +\frac{1}{2} \bar{g}_{ab} \nabla^c \nabla_c h + 2 h_{ab} - 2 h \bar{g}_{ab} = 0\,,
\end{equation}
where $h=\bar{g}^{ab}h_{ab}$. By substituting near horizon $r \ra r_+$ series expansions
\begin{equation}
    h_{ab} = e^{-i \omega v + i k \phi} (r- r_+)^\gamma \sum^{\infty}_{j=0} \tilde{h} _{ab}^{(j)} (r-r_+)^j\,,
\end{equation}
one of the Einstein equation $E_{vv}=0$ rearranges to the near horizon series form whose the leading order coefficient equation is
\begin{equation}
    \left(2\pi i T_h \omega + 4\pi i T_h \Omega k - k^2 (1 - \Omega^2) \right)\tilde{h}^{(0)}_{vv} = - (2\pi i T_h - \omega )(1 - \Omega^2)  \left[ 2 k\tilde{h}^{(0)}_{v \phi} + \omega \tilde{h}^{(0)}_{\phi \phi } \right].
\end{equation}
By that the all the coefficients of $\tilde{h}_{ab}$ vanish, the leading pole-skipping points are found as
\begin{equation}
    (\omega, k) = \left(2\pi i T_h, \pm \frac{2\pi i T_h}{1 \mp \Omega} \right)\,.
\end{equation}
This point living in incoming Eddington-Finkelstein coordinates can be transformed to the Schwarzschild coordinates by the relation \eqref{eq:wk To E-F}. The leading pole-skipping points become
\begin{equation}
    (\omega, k) = \frac{2\pi i T_h}{1-\Omega^2}\left(1\pm \Omega, \Omega \pm 1 \right)\,.
\end{equation}
This result is consistent with the conclusion \eqref{eq:Conclusion 1} for $\nu = 1$. When the conformal dimension is given by $\Delta_+ = 2$, the parameter $\nu$ becomes $1$ by the relation such as $\Delta_+ = 1 + \nu$.

The conformal dimension $\Delta_+$ is found by analyzing the holomorphic and anti-holomorphic two point functions. For the stress tensor $T$, the two point functions on the complex plane are given by
\begin{eqnarray}
\langle T(z) T(z')\rangle=\frac{c_L/2}{(z-z')^4}\,,~~~~\langle \bar{T}(\bar{z})\bar{T}(\bar{z}')\rangle=
\frac{c_R/2}{(\bar{z}-\bar{z}')^4}\,,
\end{eqnarray}
where $c_L$ and $c_R$ are the left/right central charges \cite{Liu:2020yaf}. Therefore, the conformal dimension is found as $\Delta_+=2$, because generally the two point function for an operator $D$ has a form as
\begin{equation}
    \langle D(z) D(0)\rangle \sim \frac{1}{z^{2\Delta}}\,,
\end{equation}
where $\Delta$ is a conformal dimension.


\bibliographystyle{JHEP}

\providecommand{\href}[2]{#2}\begingroup\raggedright\endgroup

\end{document}